\documentclass[manuscript, screen, nonacm]{config/acmart}

\usepackage{iftex}
\ifXeTeX
  \usepackage{xeCJK}
\fi

\usepackage{multirow}
\usepackage{subcaption}

\AtBeginDocument{%
  }

\newcommand{\Q}[1]{``#1''}
\newcommand{\projectname}{\textit{Peers4U}}

\newcommand{\paperacknowledgments}{}

\setcopyright{none}
\acmDOI{}
\acmISBN{}
\acmConference{}{}{}
\acmYear{}
\copyrightyear{}
\author{Zhi Zheng}
\affiliation{
  \institution{Department of Computer Science and Technology, Tsinghua University}
  \city{Beijing}
  \country{China}
}

\author{Binglin Liu}
\affiliation{
  \institution{Department of Computer Science and Technology, Tsinghua University}
  \city{Beijing}
  \country{China}
}

\author{Weihao Chen}
\affiliation{
  \institution{Department of Computer Science and Technology, Tsinghua University}
  \city{Beijing}
  \country{China}
}

\author{Lan Luo}
\affiliation{
  \institution{The Hong Kong University of Science and Technology (Guangzhou)}
  \city{Guangzhou}
  \country{China}
}

\author{Yu Mei}
\affiliation{
  \institution{Department of Computer Science and Technology, Tsinghua University}
  \city{Beijing}
  \country{China}
}

\author{Jie Cai}
\affiliation{
  \institution{College of Software, Nankai University}
  \city{Tianjin}
  \country{China}
}

\author{Chun Yu}
\affiliation{
  \institution{Department of Computer Science and Technology, Tsinghua University}
  \city{Beijing}
  \country{China}
}

\author{Yuanchun Shi}
\affiliation{
  \institution{Department of Computer Science and Technology, Tsinghua University}
  \city{Beijing}
  \country{China}
}

\renewcommand{\shortauthors}{Zheng et al.}

\renewcommand{\paperacknowledgments}{
}

\begin{document}
\title[Peers4U]{Peers4U: Designing Learner-Centered Virtual Peers for Children's Multimodal Mathematics Learning}

\begin{abstract}

In early mathematics, children learn abstract concepts through concrete representations, such as manipulating objects, drawing, and speaking. Learning with peers can enrich this process by exposing children to diverse strategies and enabling them to exchange ideas through interaction. We realized this potential with virtual peers in \projectname{}, a multimodal learning system to enhance children's mathematical conceptual understanding. Virtual peers driven by LLMs are personalized to the learner's cognitive level and assigned multimodal solutions with inspectable processes. During learning, peers engage in mixed-initiative interactions by responding, initiating topics, and advancing discussions along a pedagogical progression. A user study with 22 children and two experts showed our learner-centered topic orchestration significantly outperformed a persona-centered baseline, and children spontaneously authored new artifacts to demonstrate inspired ideas or share their own strategies, producing significantly more solutions when grouped with more Leading Peers. We contribute design insights for multimodal, learner-centered virtual peer systems.

\end{abstract}



\begin{teaserfigure}
  \centering
  \includegraphics[width=0.95\textwidth]{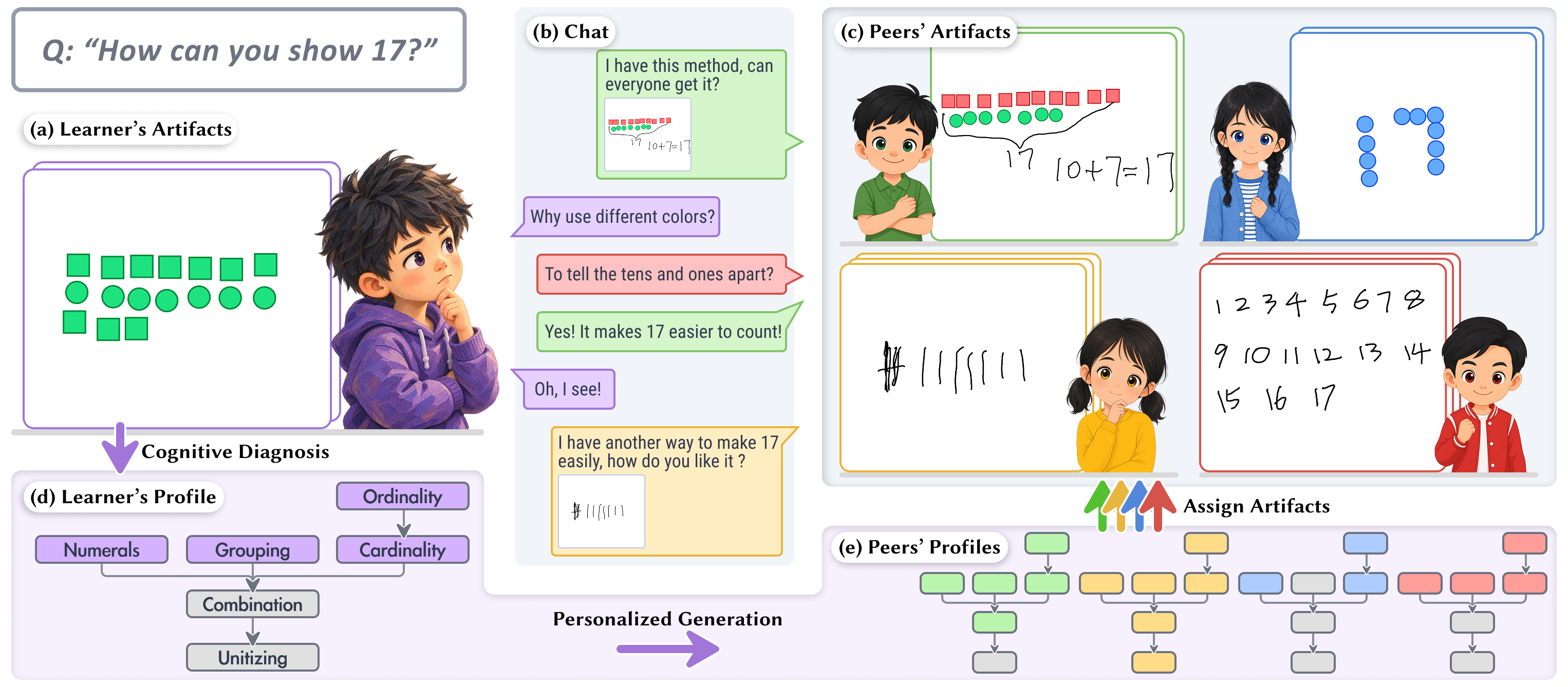}
  \caption{Overview of \projectname{} through an example task to represent 17. After authoring solution artifacts (a), the learner joins a personalized group of four virtual peers with diverse artifacts (c) and engages in discussion (b).
  The lower part of the figure shows the backend pipeline: the system analyzes the learner's initial solutions to derive a cognitive profile (d), generates heterogeneous peer cognitive profiles tailored to the learner (e), and assigns artifacts accordingly.}
  \Description{A wide composite diagram centered on the task, How can you show 17? On the left, a learner avatar accompanies an artifact made of green squares and circles; a downward arrow connects it to a task-specific cognitive graph. In the center, colored chat bubbles show the learner and peers discussing why different colors distinguish tens and ones, with artifact thumbnails embedded in two messages. On the right, four peer avatars accompany contrasting artifacts: grouped red and green shapes with the equation 10 plus 7 equals 17, blue dots arranged as the numeral 17, tally marks, and handwritten numerals from 1 to 17. Along the bottom, four simplified cognitive graphs represent the peer profiles, with arrows linking those profiles to the assigned artifacts.}
  \label{fig:0-teaser}
\end{teaserfigure}


\maketitle

\section{Introduction}

In early mathematics, it is crucial for children to ground their understanding of abstract concepts and operations in concrete experiences~\cite{baroody2017Use,ottmar2019Embodied}. For example, children can conceptualize place value by bundling ten individual objects into a group, drawing corresponding quantities, and connecting these actions to written numerals. Such multimodal learning through multiple representations, including manipulating, drawing, writing, and speaking~\cite{bruner1964Course,bruner1974Theory,leong2015ConcretePictorialAbstract,lesh1987Representations}, helps children develop deeper conceptual understanding and representational fluency~\cite{nathan2002Representational,fonger2019Meaningfulness}.

Learning with peers offers natural opportunities to enrich this process~\cite{damon1984Peer,tenenbaum2020How}. Peers often bring diverse solution strategies across different representations~\cite{zheng2026OpenCD}, exposing one another to unfamiliar ways of thinking and sparking curiosity~\cite{durkin2017Using}. Through interaction, they can further explain their reasoning, compare alternative approaches, exchange ideas, and challenge or build on one another's understanding~\cite{forman1998Exploring,schwartz1995Emergence}.

The emergence of Large Language Models (LLMs) has opened new possibilities for introducing virtual peers into Intelligent Tutoring Systems (ITSs). However, many recent virtual-peer systems target older learners or other learning domains~\cite{liu2024PeerGPT,viberg2026Codesigning,wang2025Generative}, with peer interactions primarily centered on text- and voice-based conversation~\cite{wang2026ArgueMate,yue2025MathVC,tokida2026Design}. This conversational paradigm fits poorly with the visual and manipulable needs of early mathematics learning~\cite{lesh1987Representations,ng2016Interplay}. Furthermore, little is known about how young children would participate and learn in an open virtual-peer environment.

To bridge this gap, we explore how to design a multimodal virtual peer-learning system for early mathematics. Specifically, we investigate: \textit{How can virtual peers be designed to support children's multimodal mathematics learning, and how do children participate in the resulting peer-learning interactions?}

We first conducted a formative study consisting of exploratory probe sessions with 8 children and interviews with 5 experienced teachers. We identified four interrelated design challenges: (1) children's mathematical peer communication needs to be grounded in shareable multimodal solution artifacts; (2) peers need to be sufficiently different from the learner to provide new perspectives while remaining understandable; (3) children vary substantially in how they prefer to participate; and (4) open peer interaction alone does not naturally produce a coherent learning progression. These findings informed four corresponding design goals (DGs). 

Guided by these goals, we designed and developed \projectname{}, a multimodal virtual-peer ITS for early mathematics learning, as shown in Figure~\ref{fig:0-teaser}. In \projectname{}, after independently completing multimodal solutions to an open-ended problem, the learner is presented with a personalized learning group of 4 peers based on the learner's demonstrated mathematical understanding. Each peer possesses a diverse set of solution artifacts whose multimodal construction processes can be inspected. The interaction follows a mixed-initiative design~\cite{horvitz1999Principles}: the learner can engage with peers to discuss specific artifacts and author and share new solutions at any time, while virtual peers can likewise respond, proactively initiate topics, and advance discussions along a pedagogical trajectory.

To characterize children's peer-learning experiences and evaluate the system design, we conducted an exploratory user study with 22 children across different peer-composition conditions and a post-hoc evaluation with 2 experts. \projectname{}'s learner-centered topic orchestration received high pedagogical appropriateness ratings (4.69/5), significantly outperforming a persona-centered baseline. Furthermore, children spontaneously authored new artifacts to teach peers, demonstrate newly inspired ideas, or improve and share their own strategies. Children produced significantly more additional solutions when learning with a Leading Group, while those in a Mixed Group showed significantly greater willingness to learn with the same peers again. We also identified how different types of peer messages were associated with children's likelihood of responding. We further discuss broader implications for designing multimodal virtual peers that children can both learn from and teach.

Our contributions are threefold:
\begin{itemize}
    \item We identify core challenges of introducing virtual peers into children's mathematics learning and derive four DGs spanning modality mediation, peer heterogeneity, flexible initiative, and progressive orchestration.
    \item We present the design and implementation of \projectname{}, instantiating an artifact-mediated multi-peer learning paradigm that combines mixed-initiative interaction with learner-centered pedagogical orchestration.
    \item We provide empirical findings revealing children's interaction patterns in multimodal virtual peer learning, and outline transferable design implications for virtual peer learning systems.
\end{itemize}

\section{Related Work}

\subsection{Theoretical Foundations: Multiple Representations and Peer Learning in Mathematics}

Mathematics learning is fundamentally a process of cognitive development. Rooted in Embodied Cognition, children construct abstract concepts through physical manipulation and visual perception~\cite{barsalou2008Grounded,ottmar2019Embodied}. Theories such as Bruner's Enactive-Iconic-Symbolic (EIS) framework~\cite{bruner1964Course,bruner1974Theory} and Lesh's translation model~\cite{lesh1987Representations} emphasize the necessity of practicing and translating across multiple representations---including concrete manipulatives, pictorial drawings, spoken language, and written symbols~\cite{ng2016Interplay,pape2001Role}. Engaging in these translational activities allows children to develop Representational Fluency and achieve a deep conceptual understanding~\cite{nathan2002Representational,fonger2019Meaningfulness}. Consequently, multimodal interactive learning systems inherently align with this pedagogical need, which forms the foundational rationale of our system.

Social learning environments naturally complement this need for multi-representational learning~\cite{white2011Distributed}, where peers bring diverse strategies and varying representational choices~\cite{durkin2017Using}. Drawing on Social Learning Theory~\cite{bandura1977Social}, children can effectively acquire new strategies by observing and imitating peer behaviors. Furthermore, exposure to such heterogeneous peer strategies often triggers Socio-Cognitive Conflict, compelling children to reflect on their own reasoning and ultimately facilitating deeper knowledge internalization~\cite{damon1984Peer,piaget2015Grasp}.

Consequently, peer learning paradigms are widely adopted in pedagogical practice~\cite{topping2005Trends,vass2010Peer,cooper1999Classroom}. Specifically, Reciprocal Peer Tutoring---where children alternate between tutor and tutee roles---yields mutual learning gains through preparation and explanation~\cite{fantuzzo1998Reciprocal,duran2005Styles,topping1996Effectiveness}, while cooperative learning enhances academic achievement and strengthens social support~\cite{johnson2009Educational,howe2009Peer}. Recently, advancements in Large Language Models (LLMs) have made virtual peers a reality, providing a unique opportunity to offer highly controlled and tailored peer interactions. Building on this, our research explores how to effectively design such virtual peers to facilitate multimodal mathematics learning for children.

\subsection{Multimodal Interactive Systems for Mathematics Learning}

Grounded in the theories outlined above, mathematical understanding hinges on multiple representations. Existing HCI research seeks to move beyond abstract symbols by externalizing mathematical relationships into directly perceptible, manipulable objects to scaffold learning.

One line of research combines physical environments with digital feedback to contextualize the mathematical learning experience. For instance, Tangible Tens~\cite{pontualfalcao2018Tangible} and Bricksmart~\cite{liu2025BrickSmart} leverage tabletop block play to visually represent quantitative relationships and spatial structures; Balance Board Math~\cite{tancredi2024Balance} and MOVES-NL~\cite{anton2025Mathematics,cosentino2025Exploring} map embodied action onto symbolic forms such as graphs and number lines. Extended Reality (XR) environments, such as ARMath~\cite{kang2020ARMath}, Mathland~\cite{khan2018Mathland}, and MaR-T~\cite{besevli2019MaRT}, further augment mathematical inquiry with immersive visualizations. While grounding math learning in embodied cognition provides strong intuition, these systems remain bound by the inherent properties of physical artifacts, constraining their applicability to particular mathematical topics.

Beyond physical interfaces, digital systems can cover broader learning scenarios. They can support learners' understanding of multiple interconnected representations~\cite{rau2014How,saquib2021Constructing}, using on-screen virtual manipulatives to bridge concreteness and abstraction~\cite{sarama2016Physical,moyer2002What}. Specifically, Fractions Tutor dynamically integrates graphical models with number lines~\cite{rau2009Intelligent,rau2014How}, and MathWorlds links graphs, animations, and equations~\cite{roschelle2000Simcalc}. Yet, these systems are predominantly designed for solo learning and lack built-in peer-interaction mechanisms, missing opportunities for the collision of different perspectives. While OpenCD further captures speech alongside these modalities, it is designed for diagnostic assessment rather than interactive learning~\cite{zheng2026OpenCD}.

To overcome the limitations of solo learning, we aim to introduce virtual peers embedded within a multimodal interactive environment, fostering rich cognitive friction and learning opportunities.

\subsection{Virtual Peers in Learning Environments}

Since their emergence, LLMs have been widely adopted for role-playing to generate anthropomorphic dialogue and behavior~\cite{shanahan2023Role,park2023Generative}. Building on this, educational research employs LLMs to simulate students~\cite{jin2025TeachTune,pan2025TutorUp}, laying the groundwork for virtual peers and complex social learning dynamics.

The HCI community has long focused on the design of virtual peers~\cite{cassell2000Shared, rader2011Brick,shamamyan2025CoDesigning}. Applications span from collaborative child–agent co-reading~\cite{viberg2026Codesigning} to systems like Your Thoughtful Opponent and ArgueMate, which induce cognitive conflict through disagreement and debate~\cite{kim2025Your,wang2026ArgueMate}.
Research has also explored robots as physical companions~\cite{park2017Growing,ho2021RoboMath,rutatola2026Robot,belpaeme2018Social}. A separate line of work has investigated teachable agents---where students learn by teaching AI---to enhance learning outcomes, foster metacognitive skills, and boost student engagement~\cite{xing2025Development,lyu2026Designing,jin2024Teach,sabnis2024Empowering,rogers2025Playing}.

In more social learning settings, virtual peers can participate alongside human learners~\cite{lyu2026Designinga}. For instance, PeerGPT joins children's small-group collaboration, while Saya takes part in classroom activities as an \Q{additional student}~\cite{liu2024PeerGPT,tokida2026Design}. Beyond introducing a single peer, some studies employ multiple virtual peers to create richer social learning environments, such as MAIC~\cite{yu2026MOOC}, ClassMeta~\cite{liu2024ClassMeta}, among others~\cite{yue2025MathVC,tanprasert2022Authoring,wang2025Generative,lyu2026Designinga}. However, these virtual peers primarily communicate through conversations (text/voice), which constrains their use in children's mathematics learning, where interaction often involves richer forms of engagement. Thus, how young children interact with and learn alongside virtual peers, and how such interactions can be facilitated in early mathematics, remain underexplored.

\section{Formative Study}

To inform the design of our system, we conducted a two-part formative study consisting of exploratory probe sessions with children and semi-structured interviews with elementary mathematics teachers. We synthesized evidence from both parts into four design goals (DGs).

\subsection{Study Rationale and Design}
We specifically focused on the early stages of mathematical and social learning---spanning late kindergarten to early primary school. Mathematics at this stage is strongly grounded in concrete and visual representations. Rather than relying solely on verbal explanations, children at this stage commonly express and develop mathematical ideas through manipulating objects, drawing, writing, and speaking. This setting therefore provides a particularly relevant context for examining how virtual peers might participate in and support multimodal mathematical learning.

Our formative study combined two complementary perspectives. First, we used an exploratory probe to observe how children attended to, interpreted, and interacted with peers' multimodal mathematical solutions. Second, we interviewed experienced teachers to contextualize these behaviors within classroom practice and to identify pedagogical opportunities, risks, and desirable forms of peer interaction.

\subsection{Part 1: Child Probe Sessions}

\textbf{Probe.}
We developed an exploratory probe that intentionally kept the social behaviors of virtual peers relatively simple while providing a rich and open multimodal solution space. The probe was intended to elicit how children and teachers expected peers to participate in such a space and what forms of interaction they found useful or problematic.

\begin{figure}[htbp]
  \centering
  \includegraphics[width=0.6\linewidth]{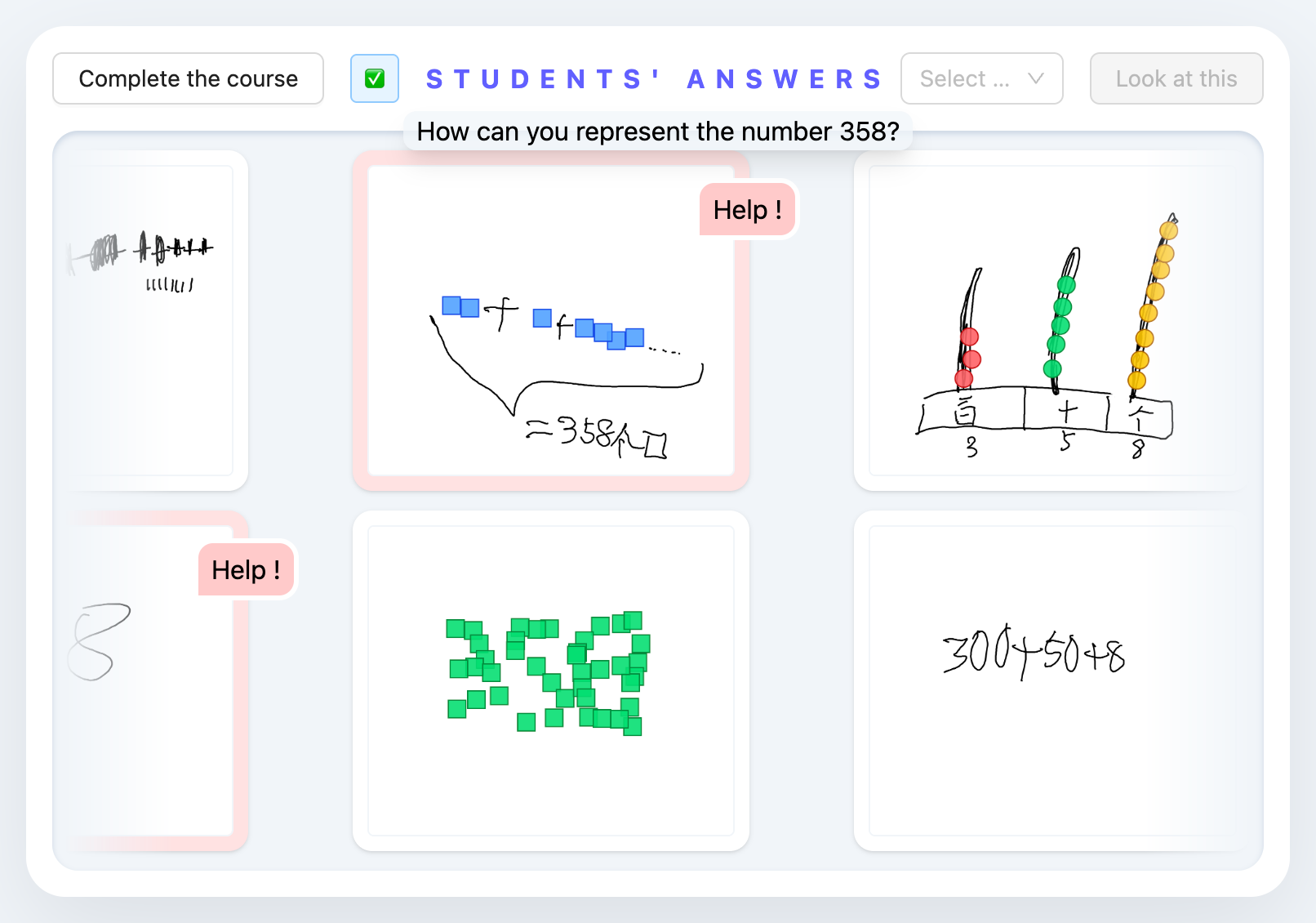}
  \caption{User interface of the probe system. Each card presents a multimodal artifact; children could replay its construction process and chat with the associated virtual peer.}
  \Description{A screenshot of the probe's peer-solution gallery for the task, How can you represent the number 358? Six cards display varied child-created representations, including tally marks, manipulatives with a handwritten equation, colored beads arranged by hundreds, tens, and ones, clusters of green squares, and the expression 300 plus 50 plus 8. Two incomplete or flawed solutions are outlined in pink and marked with Help labels.}
  \label{fig:3-probe}
\end{figure}

Children first selected a mathematics task and created one or more solutions using multiple modalities, including manipulating virtual objects, drawing, writing, and verbal explanation, using a multimodal authoring workspace described in prior work~\cite{zheng2026OpenCD}. After submitting their own solution, children were presented with a gallery of peer solutions, as shown in Figure~\ref{fig:3-probe}. They could replay the multimodal process of each solution and chat with the associated LLM-driven virtual peer via voice input and text-to-speech (TTS) output. Artifacts included flawed solutions where peers actively asked for help.

The peer-solution repository was constructed from an existing dataset described in OpenCD~\cite{zheng2026OpenCD}, with its reuse approved under the applicable IRB protocol. All data were de-identified prior to use, and audio responses additionally underwent voice conversion to reduce speaker identifiability. The mathematics tasks are listed in Appendix~\ref{apx:formativeProblems}; details of the peer-solution presentation order and the LLM configuration of the virtual peers are provided in Appendix~\ref{apx:formativeProbe}.

\textbf{Participants.}
We recruited 8 children (4 boys and 4 girls; age $M = 6.7$ years, $SD = 0.61$, range = 5y11m--7y5m) spanning the final year of kindergarten through first grade (4 K, 4 G1). Recruitment was conducted via personal accounts on RedNote for in-person sessions in a Chinese city. Parents registered their children voluntarily, and we filtered excess applicants solely to balance gender and grade. Following IRB approval and parental written consent, children participated voluntarily and were informed that they could pause or withdraw at any time. They received certificates as gifts upon completion. Participant details are provided in Appendix~\ref{apx:formativeChildren}.

\textbf{Procedure.}
Each child participated in an approximately 60-minute, in-person session using the probe on a 13-inch iPad with an Apple Pencil. The probe included nine mathematics tasks across three categories (Appendix~\ref{apx:formativeProblems}). Children could freely choose tasks, complete one or more solutions, browse peer solutions, inspect their multimodal construction processes, and interact with the corresponding virtual peers. We did not explicitly tell children beforehand that the peers were AI-driven; children could nevertheless infer this from occasional imperfections in the interactions.

The sessions followed a semi-structured observation protocol: as children navigated the system, the researcher occasionally probed their willingness to engage in specific interactions, prompted them to articulate their feelings and thoughts, and observed their spontaneous behaviors. Children could work on multiple tasks during the session, and the researcher occasionally varied the peer-solution presentation configuration to elicit user reactions across different social contexts (Appendix~\ref{apx:formativeChildren}).

\subsection{Part 2: Teacher Interviews}

\textbf{Participants.}
We recruited 5 experienced elementary school mathematics teachers from different schools through personal networks. They had 7--17 years of teaching experience ($M = 11.2$, $SD = 4.15$), including 3--17 years of experience teaching Grades 1--2 ($M = 7.2$, $SD = 5.67$). Participants received 200 RMB (approx. \$30 USD) for the 1.5-hour interview. See Appendix~\ref{apx:formativeTeachers} for detailed teacher profiles.

\textbf{Procedure.}
We conducted semi-structured interviews centered on three themes: (1) the role and dynamics of peer interaction in real-world early math classrooms; (2) their professional critique and design suggestions after the demonstration of our technology probe; and (3) their interpretations and pedagogical attributions of the specific child behaviors we observed in Part 1.

\subsection{Data Collection and Analysis}

Throughout the study, we collected screen recordings, video footage, and raw interaction logs from the children's sessions, alongside audio recordings from the teacher interviews. All audio was transcribed verbatim. 

We conducted a team-based codebook thematic analysis~\cite{braun2006Using}. The child-session and teacher-interview data were first analyzed separately. For the child-session data, screen recordings and video footage were synchronized and analyzed together, using segments containing a coherent expression or behavior as analytic units. Two researchers independently open-coded a subset of each dataset (three child sessions and two teacher transcripts) and developed codebooks. One researcher coded the remaining data, while the second reviewed the coded segments. Within each dataset, related codes were grouped into candidate themes and reviewed against the full data.

Finally, the researchers compared and integrated themes across the two datasets, connecting children's observed behaviors with teachers' pedagogical interpretations.

\subsection{Findings and Design Goals}

Overall, the probe successfully elicited rich interactions. Children selected an average of $4.0 \pm 1.2$ tasks, initiated chats around $5.5 \pm 4.1$ peer artifacts, and sent $19.6 \pm 16.5$ messages. Through our observations of children's interactions with the probe and interviews with expert teachers, we identified four core challenges and derived corresponding Design Goals (DGs) for integrating virtual peers into multimodal early math learning.

\paragraph{\textbf{\upshape Finding 1: Peer interaction around mathematics extends beyond language to multimodal artifacts}}

A key challenge is that text or speech alone could not accommodate how children wanted to communicate mathematical ideas. In our probe sessions, children often sought to share or demonstrate their own multimodal solutions, making solution artifacts part of the \Q{language} through which they observed, questioned, explained, and responded to one another.

The dynamic solution process drew particular attention, and children frequently directed their questions to specific actions within it (S2, S4, S5, S8). For example, S4 noticed the sequence of a peer's actions, asking, \Q{Why did you drag 8 squares first, and then 8 circles?} Teachers similarly emphasized the importance of making mathematical thinking visible (T1, T4). As T4 explained, young children often need to \Q{see it while listening} in order to follow an explanation.

Solution artifacts also became a natural medium for interacting with peers. When attempting to respond or explain, some children wanted to directly share or demonstrate their own solutions rather than rely on verbal explanation (S1, S2, S5). Peer solutions could also prompt further creation: S4 and S7 voluntarily proposed returning to the canvas to create another solution after viewing others' approaches. Children sometimes even used their solutions for self-presentation, proudly telling peers how many methods they had created (S4, S6).

\textbf{$\Rightarrow$ DG1: Ground peer interaction in shareable multimodal mathematical artifacts.}
Peer interactions should revolve around mathematical solutions rather than purely conversational text. Artifacts should serve not only as initial assessments but also as an ongoing medium of communication. The system should allow children to revisit, share, and discuss artifacts during interaction, as well as create new solutions.

\paragraph{\textbf{\upshape Finding 2: Useful peers must be different enough to inspire, yet close enough to understand}}

A second challenge lies in calibrating peer heterogeneity. Across all eight children, overly simple or similar solutions were often quickly skipped. Conversely, unfamiliar methods sparked curiosity, with 7 children expressing a desire to communicate upon seeing a novel solution. However, when the cognitive gap was too large, children sometimes misinterpreted the methods or incorrectly judged them as wrong (S2, S4, S5). Teachers therefore noted that peer materials must provide an \Q{intermediate ladder}, guiding children from understandable methods toward unfamiliar ones (T2, T4).

We also observed the unique pedagogical value of peers presenting less developed or flawed solutions.
Six children showed a strong willingness to help when peers explicitly asked for assistance, and re-explained their own thoughts while helping (S1--S6). S1 eagerly claimed, \Q{Of course I want to help, I'll be the little teacher!} Some children even transferred strategies they had just learned from another peer (S1, S3). Teachers noted that explaining to others can force children to reorganize their reasoning and expose gaps in their own understanding (T2, T3).

Consequently, different peer levels serve distinct purposes: more advanced peers inspire, similar peers provide relatability, and struggling peers offer opportunities for correction and confidence-building (T2--T5).

\textbf{$\Rightarrow$ DG2: Construct heterogeneous peers relative to the learner.} The system should dynamically assign a group of peers with diverse mathematical approaches and cognitive levels, tailored to the user's initial understanding. This heterogeneity must balance inspiration with approachability, deliberately providing diverse opportunities for observation, explanation and learning-by-teaching~\cite{topping1996Effectiveness,forman1998Exploring, fantuzzo1998Reciprocal}.

\paragraph{\textbf{\upshape Finding 3: Children's participation styles vary substantially, requiring flexible interaction initiative}}

A third challenge is the substantial variation in how children prefer to participate in peer learning. A rigid interaction paradigm---whether primarily system-led or learner-led---is unlikely to work well for all children.

Interaction logs revealed a participation divide. On average, some children sent over 10 messages per task (S3, S6), while others sent no more than 1 message per task (S7, S8). S7 even explicitly rejected a peer's request for help and later explained, \Q{I don't want to talk to them anymore because I want to go to the next problem faster.} T5 pointed out that while peer presence encourages some children to express themselves, it can make others more silent, thus for introverted learners, \Q{listening while other peers ask questions} was also an effective and necessary form of participation.

\textbf{$\Rightarrow$ DG3: Support flexible interaction initiative and participation roles.}

The system's conversational agency must be flexible, as reflected in prior learning systems that combine learner- and agent-initiated interaction~\cite{yu2026MOOC,wang2025Generative}. Virtual peers should be able to naturally respond to user inquiries, but also proactively initiate topics, move the discussion forward, or gracefully conclude them when the user is passive. This allows children with different participation styles to fluidly shift among the roles of leader, responder, or observer.

\paragraph{\textbf{\upshape Finding 4: Open Peer Interaction Lacks Focus Without Progressive Pedagogical Orchestration}}

Finally, young children may lack the metacognitive capacity to independently navigate multiple peers, representations, and mathematical approaches. An open peer environment does not by itself produce an effective learning trajectory. As T3 emphasized, \Q{Learning is not just about interaction\ldots{} it must have clear pedagogical objectives.}

Teachers outlined a consistent progression for productive peer learning. First, children must form their own initial ideas before being exposed to peers (T1, T5). Subsequently, peer materials should be introduced progressively---starting from relatable, easier methods to increasingly diverse representations (T1, T2, T4). Furthermore, T1 noted that after exploring divergent methods, there must be a convergent process where children are guided to compare differences, analyze commonalities, and extract mathematical essences.

This suggests a tension between \emph{social openness} and \emph{pedagogical structure}: peer interaction should feel exploratory, while its underlying progression still needs to be organized toward specific learning objectives.

\textbf{$\Rightarrow$ DG4: Orchestrate peer interaction through cognitively progressive learning topics.}
The system must act as an \Q{invisible director}, dynamically orchestrating learning topics around the learner's current understanding. The discussion of solution methods should follow the developmental trajectory of mathematical cognition~\cite{fuson1997Childrens,siegler1984Strategy}, and the types of discussion topics should deepen progressively according to cognitive levels~\cite{krathwohl2002Revision}.

\section{System Design}

In this section, we present the design of \projectname{} based on the four DGs derived from our formative study. Together, they shaped a holistic interaction paradigm for \projectname{}.

\subsection{System Overview}

For each task, the learner first independently constructs their own solutions. This serves multiple purposes: it ensures independent mathematical thinking, provides materials for subsequent peer interaction, and allows the system to estimate the learner's current mathematical understanding.

\begin{figure}[htbp]
  \centering
  \includegraphics[width=0.9\linewidth]{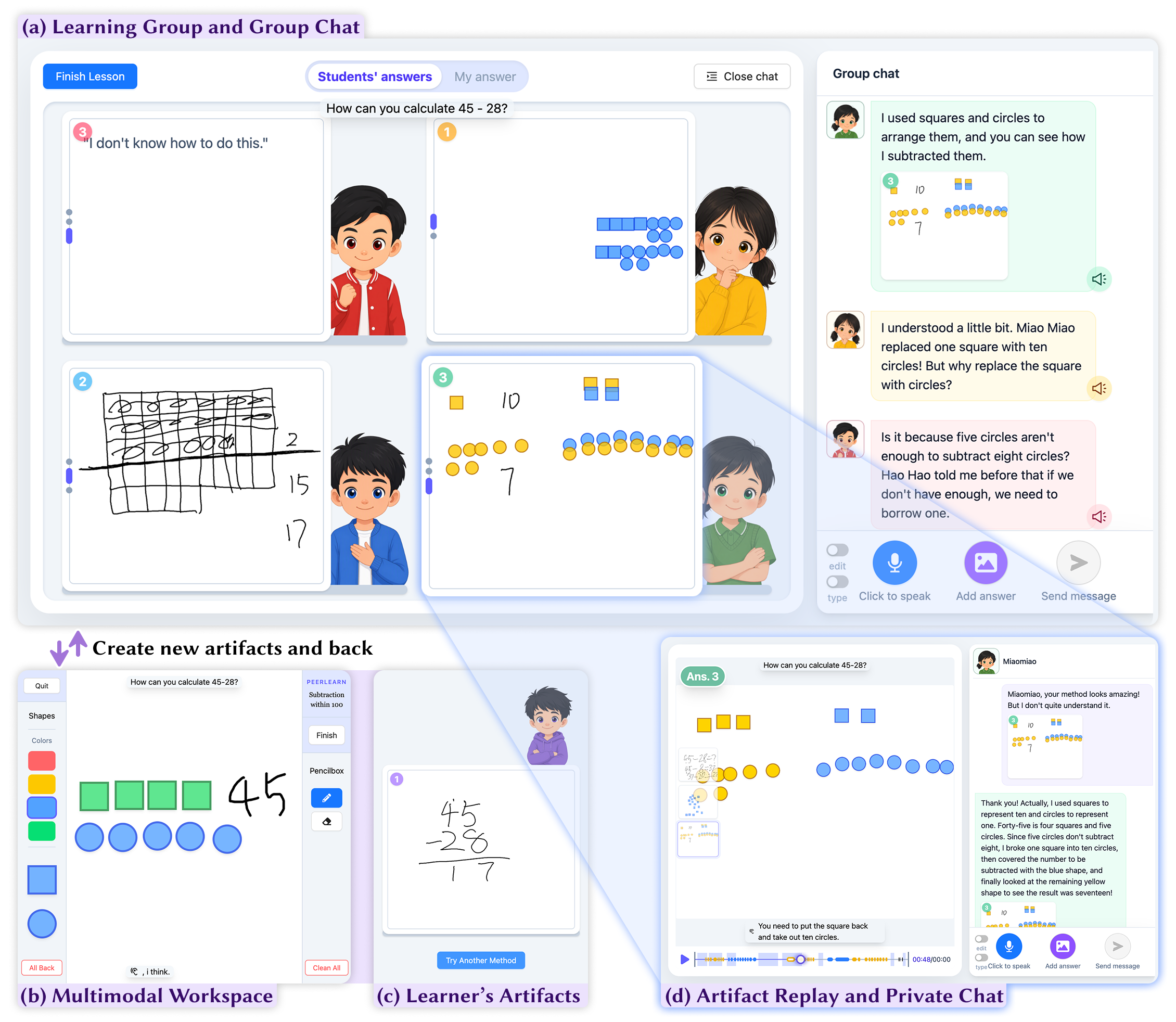}
  \caption{User interface and interaction flow of \projectname{}. Learners can browse all peers' artifacts and chat with the group (a). They can also create new artifacts in the multimodal workspace (b), revisit their own artifacts (c), or replay a peer's multimodal solution process and discuss with them in a private chat (d).}
  \Description{A composite of four connected interface views. The large upper view shows four virtual peers and their artifacts in a two-by-two learning-group grid, with a group-chat panel containing text, audio controls, and shared artifact thumbnails. A translucent arrow connects this view to a multimodal workspace at the lower left, where the learner can draw, write, speak, and manipulate colored shapes. The lower middle view shows the learner's saved artifacts. The lower-right view combines a peer artifact replay with a playback timeline and a private chat grounded in the displayed artifact.}
  \label{fig:4-UI}
\end{figure}

Following this, the learner enters a virtual \Q{classroom}, as shown in Figure~\ref{fig:4-UI}, where the system has tailored a personalized learning group (DG2). Mirroring real-world small-group dynamics, this group consists of four virtual peers, each equipped with multiple multimodal solution artifacts. Learners can review their peers' problem-solving processes, engage in group or private chats, directly discuss specific artifacts, and create and share new artifacts at any time (DG1). Furthermore, peers can proactively initiate learning topics to engage the user and move the discussion forward (DG3). These topics are diverse, unfold iteratively, and are dynamically orchestrated based on the learner's performance, progressing from shallow to deep following pedagogical strategies (DG4).

Ultimately, these mechanisms aim to foster deep participation, effective peer collisions, and an orderly deepening of conceptual understanding. Below, we detail the system's design.

\subsection{Multimodal Artifact Creation and Peer Interaction (DG1)}

DG1 requires the interface to facilitate natural, multimodal interactions centered around mathematical artifacts.

To support children's mathematical expression and learning through multiple representations, \projectname{} allows learners to create solutions using a variety of modalities, including virtual manipulatives~\cite{sarama2016Physical,moyer2002What}, drawing, writing, and speaking, as shown in Figure~\ref{fig:4-UI}b. We adopted the multimodal workspace introduced in OpenCD~\cite{zheng2026OpenCD} for this purpose.

After completing their initial solutions, learners enter the \Q{Learning Group} page (Figure~\ref{fig:4-UI}a). The four peers and their associated solutions are presented together for an overview. Learners can open the chat panel to participate in group discussions, or inspect a peer's solutions, replay their complete multimodal construction process, and communicate privately with that peer (Figure~\ref{fig:4-UI}d).

Beyond text or voice chat, students can return to the workspace at any point to author new content, and send their entire multimodal creation process to peers. Naturally, students can also forward other peers' solutions into the chat for discussion, a capability shared by the virtual peers as well.

\subsection{Personalized Peer Construction (DG2)}

DG2 requires the system to dynamically assemble an appropriate peer group for each learner. Specifically, the system needs to estimate the learner's current mathematical understanding, determine heterogeneous but appropriate cognitive profiles for multiple peers, and assign each peer a set of solutions consistent with that profile.

\subsubsection{Artifact Analysis and Cognitive Mapping}
\label{sec:artifactAnalysis}

A learner's cognitive level can be estimated from their initial solution artifacts. The system must comprehend the full multimodal problem-solving process and map it to task-specific mathematical competencies. Because \projectname{} uses the same mathematical tasks and multimodal workspace as OpenCD~\cite{zheng2026OpenCD}, we directly reused its knowledge bases, which contain typical solution behaviors and task-specific cognitive graphs organized as directed acyclic graphs (DAGs), and adapted its VLM-based artifact-analysis pipeline:

\textbf{Solution Artifact Comprehension Agent:} Translates the multimodal problem-solving process into a concise text description, enabling downstream interaction agents to efficiently handle a large volume of artifacts during conversations.

\textbf{Behavior Detection and Cognitive Mapping Agent:} Identifies task-relevant behavioral evidence from the artifact and maps it to corresponding nodes in the cognitive graph. This indexes the artifact at both behavioral and cognitive levels, establishing a structural correspondence between the solutions and cognitive states.

\begin{figure}[htbp]
  \centering
  \includegraphics[width=0.8\linewidth]{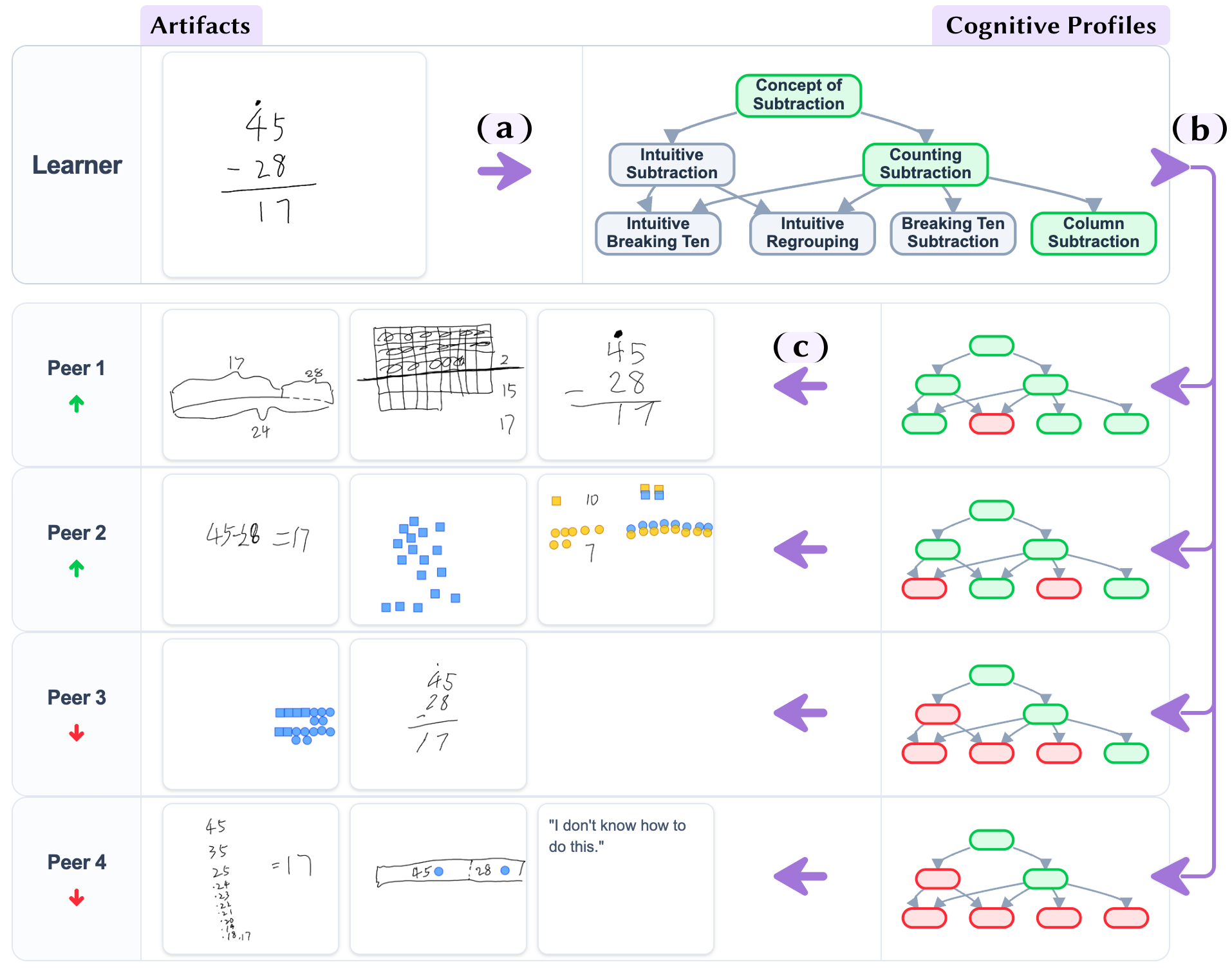}
  \caption{Example of personalized peer construction. The system maps the learner's artifacts to a task-specific cognitive profile (a), constructs heterogeneous profiles for four peers (b), and assigns each peer a diverse portfolio of artifacts consistent with their profile (c). Within the profiles, green and red nodes denote mastered and unmastered competencies, respectively, while gray nodes denote competencies not demonstrated. Green upward and red downward arrows on the left indicate Leading and Struggling Peers, respectively.}
  \Description{A three-stage flow diagram arranged in one learner row followed by four peer rows. The learner's vertical-subtraction artifact for 45 minus 28 points to a labeled cognitive graph whose nodes are green or gray. The graph then branches into four simplified peer profiles: two Leading Peers marked by green upward arrows and two Struggling Peers marked by red downward arrows. Arrows from these profiles point to portfolios of two or three artifacts per peer. The Leading Peers receive varied, mostly complete representations, while the Struggling Peers receive less developed, incomplete, or flawed representations.}
  \label{fig:4-peers}
\end{figure}

Consequently, a student's initial solution artifacts are mapped as a set of evidence on the cognitive graph, as shown in Figure~\ref{fig:4-peers}a, forming a task-specific cognitive profile that supports subsequent peer construction.

Furthermore, all artifacts within the system's artifact repository are preprocessed using the same pipeline, making them ready for peer allocation in the subsequent step.

\subsubsection{Assigning Artifacts to Peers}

Based on the learner's cognitive profile and initial solutions, \projectname{} constructs multiple peers with heterogeneous cognitive levels and assigns each peer a diverse set of appropriate artifacts. Leveraging the aforementioned analysis preprocessing, this construction can be performed using explicit rules rather than LLM generation, making the process more efficient and controllable.

Following DG2, we distinguish two broad peer roles relative to the learner: \emph{Leading Peers} and \emph{Struggling Peers}. Leading Peers are assigned valuable strategies that the learner has not yet demonstrated, providing opportunities to encounter new perspectives. Struggling Peers contain more incomplete or flawed solutions, creating opportunities for correction and learning-by-teaching. Peers within the same role can also differ in their specific cognitive profiles. The system supports flexible combinations of these roles; the configurations used in our user study are described in \S~\ref{sec:userStudySetup}.

Once the peers' cognitive profiles are determined (Figure~\ref{fig:4-peers}b), the system assigns solutions that both support these personas and provide useful materials for interaction (Figure~\ref{fig:4-peers}c). Motivated by prior work showing both the benefits and cognitive demands of learning from multiple solution methods~\cite{grosse2006Effects,rittle-johnson2007Does}, we bounded each group's solution pool around 10 artifacts, with at least 2 per peer. Using the preprocessing results described above, the assignment procedure prioritizes solutions according to their cognitive fit with the target peer, similarity to existing solutions and modality coverage. This produces peer portfolios that are consistent with the intended cognitive profiles while remaining diverse in strategy and form. The construction and assignment rules, together with examples across tasks and group compositions, are provided in Appendix~\ref{apx:personalizedPeerConstruction}.

\subsection{Learning Topic Initiation and Progression (DG3, DG4)}

Rather than allowing each virtual peer to independently pursue a persona-driven agenda, \projectname{} coordinates peer behaviors around the learner's learning progression to achieve learner-centered interaction. To fulfill this, we design the interaction mechanism from two dimensions: adaptive conversational agency and progressive topic orchestration.

\subsubsection{Adaptive Conversational Agency via Implicit Interaction}

DG3 requires the coordination to remain flexible: peers should respond when the learner takes initiative, while also being able to proactively drive the discussion when appropriate, consistent with mixed-initiative interaction principles~\cite{horvitz1999Principles}. However, rigid system interventions can easily interrupt the child's active learning process. 
We operationalize this through \textbf{implicit interaction}: using the user's operational silence as an implicit contextual cue to dynamically modulate conversational agency~\cite{serim2019Explicating,chen2026When}.

\begin{figure}[htbp]
  \centering
  \includegraphics[width=0.8\linewidth]{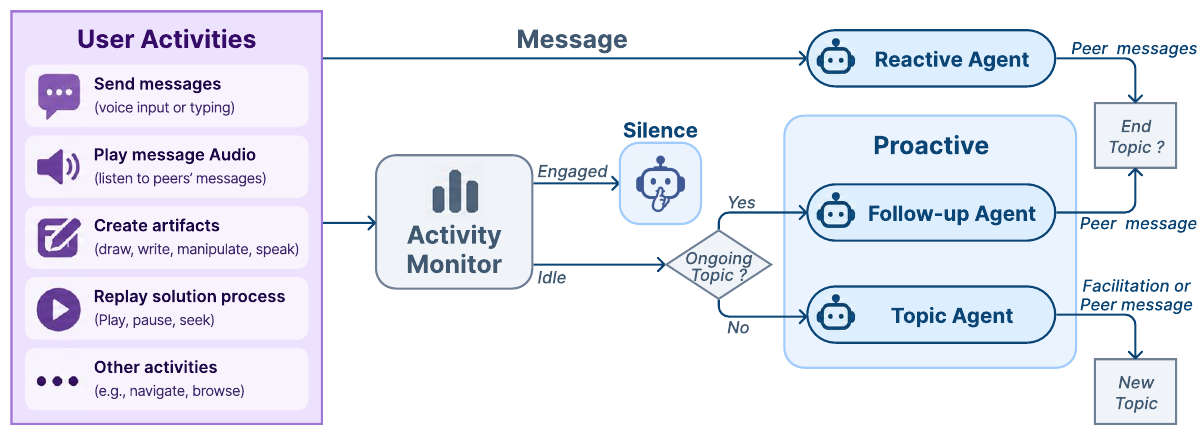}
  \caption{Message-generation flow of \projectname{}. A learner message invokes the Reactive Agent directly. Other activities are monitored to estimate the learner's state. If the learner is engaged, the system remains silent; if idle, it invokes either the Follow-up Agent or the Topic Agent. Both the Reactive and Follow-up Agents can conclude the current topic.}
  \Description{A left-to-right flowchart. A list of learner activities includes sending messages, playing message audio, creating artifacts, replaying solution processes, and navigating. Messages go directly to the Reactive Agent. All activities also enter an Activity Monitor: an engaged state leads to Silence, while an idle state leads to a decision about whether a topic is ongoing. An ongoing topic invokes the Follow-up Agent; otherwise, the Topic Agent initiates a new topic. The Reactive and Follow-up Agents produce peer messages and may end the current topic, whereas the Topic Agent produces a facilitation or peer message for a new topic.}
  \label{fig:4-agents}
\end{figure}

The system continuously monitors the learner's activity stream and triggers generation through a conversation control flow shown in Figure~\ref{fig:4-agents}. When the learner sends a chat message, the system directly invokes the \textit{Reactive Agent} for a response. For other activities, the system differentiates the user's state: if the user is engaged in high-initiative tasks such as recording speech, constructing an artifact, or replaying a solution process, the system classifies the state as \textit{Engaged} and remains silent to avoid interruption. Conversely, after the learner remains \textit{Idle} for a period, the system treats this as an opportunity for a lightweight proactive intervention. It invokes the \textit{Topic Agent} to initiate a new topic or the \textit{Follow-up Agent} to advance the current discussion. Both the Reactive and Follow-up Agents can close the current topic when the topic's goal has been reached or has expired. Implementation details are provided in Appendix~\ref{apx:adaptiveConversationalAgency}.

Through this mechanism, children who prefer to lead can do so without interruption, while those who participate less actively can still receive timely invitations to engage, fostering an elastic and natural group chat rhythm.

\subsubsection{Orchestrating Progressive Learning Topics}

To ensure an orderly deepening of discussions (DG4), we designed various topic types incorporating suggestions from our formative study teachers, roughly corresponding to different levels of Bloom's Taxonomy~\cite{krathwohl2002Revision}. Because some advanced topics require pedagogical orchestration rather than a peer's perspective, we introduced \textit{Facilitation Messages}. These act similarly to a teacher's guiding prompts or group announcements, injected into the chat to stimulate peer-wide discussion. The topic types are detailed in Table~\ref{tab:topics}.

\begin{table*}[ht]
\centering
\small
\caption{Taxonomy and types of learning topics.}
\Description{A three-column taxonomy table linking four levels of learning objectives to six topic types and their purposes. Understanding includes peer-led sharing and verification and clarification requests; applying includes peer-led help requests; analyzing includes facilitation-led prompts to find commonalities and compare differences; and evaluating includes a facilitation-led overall evaluation.}
\label{tab:topics}
\begin{tabular}{p{0.12\linewidth} p{0.19\linewidth} p{0.61\linewidth}}
\toprule
\textbf{Taxonomy} & \textbf{Topic Type} & \textbf{Description} \\
\midrule
\multirow{2}{*}{Understanding} 
& Sharing and Verification \newline \textit{(Peer Message)} & A peer shares their artifact in the group and asks if others understand, introducing new strategies and perspectives. \\
\cmidrule{2-3}
& Clarification Request \newline \textit{(Peer Message)} & A peer privately asks about the learner's artifact and invites the learner to explain its meaning or reasoning, encouraging them to externalize their understanding. \\
\midrule
Applying 
& Help Request \newline \textit{(Peer Message)} & A peer privately seeks help with a flawed artifact, asking the learner to identify errors or demonstrate fixes, thereby creating a chance for learning-by-teaching. \\
\midrule
\multirow{2}{*}{Analyzing} 
& Finding Commonalities \newline \textit{(Facilitation Message)} & Invites the group to analyze commonalities between two artifacts, converging from surface forms to mathematical essence. \\
\cmidrule{2-3}
& Comparing Differences \newline \textit{(Facilitation Message)} & Invites the group to identify differences between two distinct artifacts and evaluate advantages, establishing connections between alternatives. \\
\midrule
Evaluating 
& Overall Evaluation \newline \textit{(Facilitation Message)} & Invites the group to review all solutions and evaluate the strengths and limitations, and encourages creating new artifacts, fostering knowledge internalization. \\
\bottomrule
\end{tabular}
\end{table*}

The Topic Agent leverages these topic types to dynamically structure the interaction over time. Its prompt encodes the intended pedagogical progression, enabling the agent to select the next topic based on the learner's demonstrated understanding, peers' cognitive configurations, and prior chat history. Guided by this logic, earlier interactions primarily draw on lower-demand peer messages, which provide opportunities to practice mathematical explanation and probe the learner's understanding. As interaction progresses, deeper analytical topics are introduced to connect multiple solutions and reveal relationships among mathematical strategies. Finally, an evaluation topic is deployed as a concluding discussion to scaffold convergence and bring the learning sequence to a coherent close.

\section{User Study}

We conducted a user study to answer the following research questions:

\textbf{RQ1:} How do children engage in and experience multimodal peer learning with \projectname{}?

\textbf{RQ2:} How does peer composition influence children's engagement and experience?

\textbf{RQ3:} How pedagogically appropriate are the learning topics generated from children's actual interactions?

We conducted an exploratory, in-person, between-subjects study with children to address RQ1 and RQ2, and a post-hoc evaluation with two expert teachers to answer RQ3.

\subsection{Children Participants}

We initially recruited 24 children, while two withdrew, and their incomplete records were excluded. The final sample comprised 22 children (11 girls and 11 boys; age $M=7.00$ years, $SD=0.67$, range = 6y0m--8y2m, denoted as U1--U22). The study took place during the summer break; 11 children had completed their final kindergarten year and 11 had completed Grade~1. Participant details are provided in Appendix~\ref{apx:userStudyParticipants}.

Recruitment was conducted via personal accounts on RedNote for in-person sessions in a Chinese city, with parents registering their children voluntarily. Because learning ahead of the school curriculum is common in this context, parents reported their child's actual mathematics progress. We excluded applicants who had already completed the Grade~2 curriculum to minimize ceiling effects, and filtered excess applicants solely to balance gender and grade. Following IRB approval and parental written consent, children participated voluntarily and could pause or withdraw at any time. Each child received a small toy and a certificate as gifts upon completion.

\subsection{Study Setup}
\label{sec:userStudySetup}

We employed a between-subjects design comparing two configurations ($n=11$ each): \textbf{Condition A (Mixed Group)} included two Leading and two Struggling Peers, providing a balanced cognitive distribution; \textbf{Condition B (Leading Group)} included four Leading Peers, maximizing exposure to inspiring strategies. In both conditions, peer cognitive profiles and solution artifacts were personalized based on the learner's initial solutions. Peer appearance, screen position, and voice were randomized for each participant, with two male and two female peer characters per group.

\begin{figure}[htbp]
  \centering
  \includegraphics[width=0.4\linewidth]{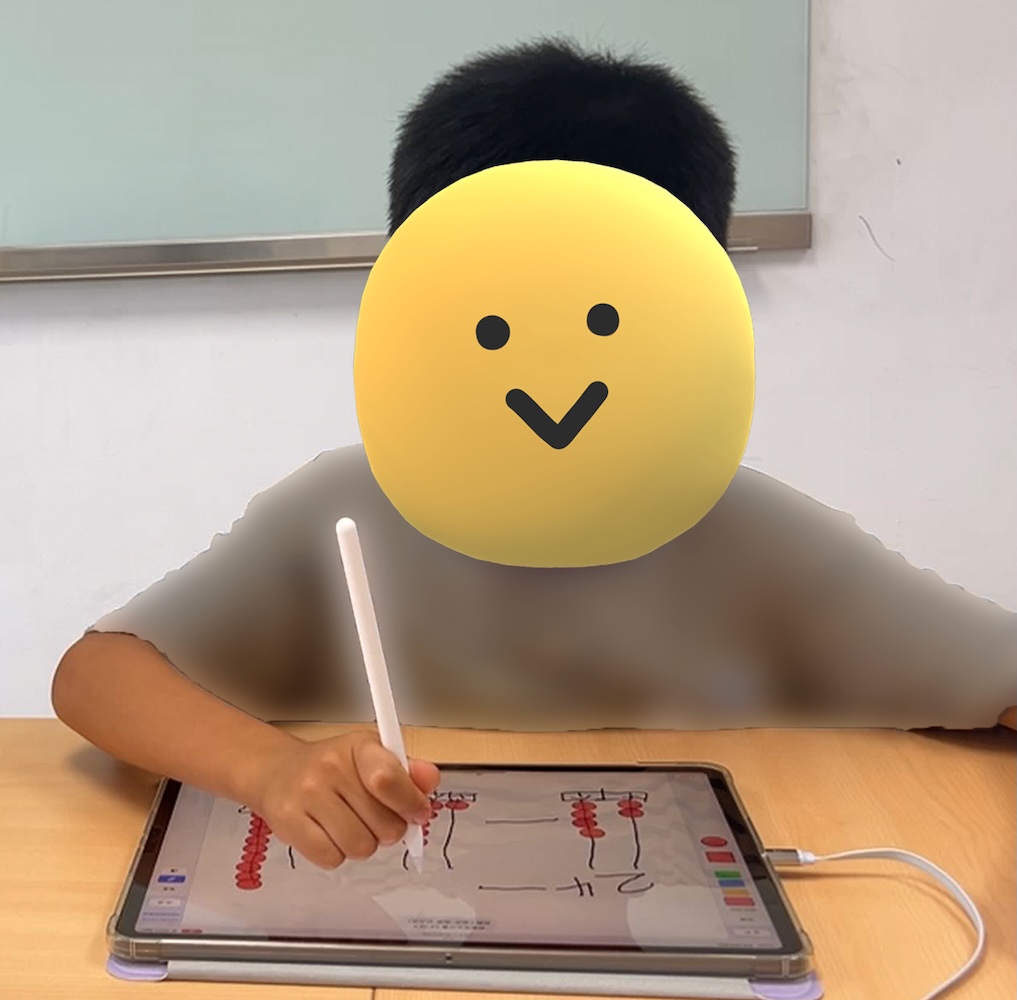}
  \caption{In-person user-study setup. A child used \projectname{} on a 13-inch iPad Pro with an Apple Pencil.}
  \Description{A photograph of a child seated at a desk and using a white stylus on a tablet. The tablet displays the multimodal mathematics workspace with red virtual manipulatives and handwritten marks. The child's face is covered by a smiley graphic and the upper body is blurred to obscure identifying features; a classroom whiteboard appears in the background.}
  \label{fig:5-setup}
\end{figure}

All sessions used a 13-inch iPad Pro with an Apple Pencil, as shown in Figure~\ref{fig:5-setup}. We selected \textit{Subtraction with Regrouping} as the learning task and prepared three task tiers corresponding approximately to entry into Grade~1, the first semester, and the second semester. Each tier paired a focal learning task with a different but comparable post-test task (Appendix~\ref{apx:userStudyProblems}). The artifact repository for these tasks was assembled from the OpenCD dataset~\cite{zheng2026OpenCD}, our formative study, and researcher-authored supplements, with details provided in Appendix~\ref{apx:userStudyProblems}; all artifacts were preprocessed using the pipeline described in \S~\ref{sec:artifactAnalysis}.

\subsection{Procedure}

Each child first completed a guided tutorial covering the multimodal workspace, solution replay, and chats. Then the researcher suggested a task tier based on the child's grade and parent-reported progress; the child could confirm or adjust this choice. Participants were then assigned to one of the two conditions to balance task tier and gender.

Children then completed one full learning session in \projectname{}. Their independently authored initial solutions served as the pre-test and as the input for personalizing the four peers. Subsequently, they explored peer solutions and interacted with the peers. The system explicitly notified the user when planned learning topics were completed, though children were free to continue interacting if they wished. Children could also choose to end the lesson or decline any remaining procedures at any time. After the lesson, children completed a questionnaire, which provided a short break from problem solving and then authored solutions to the paired post-test task. A brief structured interview concluded the session.

The researcher remained present during the formal task, but assistance was strictly limited to reading text for children who needed it, correcting speech-recognition transcripts, and answering basic interface queries.

\subsection{Data Collection and Analysis}

We collected camera and screen recordings, along with system logs capturing the complete pre- and post-test solution processes, additional solutions created during learning, chat messages, and interaction events. The post-study questionnaire captured perceptions of individual peers, social pressure, and comparisons with regular classes. Most ratings used a five-point pictorial Smileyometer~\cite{hall2016Five}. The interview probed perceived learning, preferred solution strategies, experiences of giving and seeking help, and liked or disliked aspects of the experience. The complete instruments are provided in Appendix~\ref{apx:userStudyInstruments}. 

We combined descriptive analyses of logs and questionnaires with structured coding of messages, multimodal artifacts, and interviews. Messages and artifacts were coded using researcher-defined schemes, and structured interview responses were analyzed question by question using qualitative content analysis. Between-condition outcomes were compared using two-sided exact Mann--Whitney $U$ tests, while message-level response comparisons used user-stratified exact permutation tests. Coding procedures and statistical details are provided in Appendices~\ref{apx:userStudyAnalysis} and~\ref{apx:resultStatistics}.

\subsection{Post-hoc Expert Evaluation}

To assess the pedagogical quality of system-initiated topics, we conducted an expert evaluation examining (1) whether topic messages were pedagogically appropriate, (2) whether our learner-centered orchestration produced more appropriate topics than a persona-centered agent, and (3) why topics were sometimes inappropriate.

\textit{Evaluation Dataset.}
We evaluated all 157 topic message points from U1--U22. Each paired the learner-centered message delivered during the user study with a persona-centered counterpart generated from the same preceding interaction context. The latter initiated topics from a peer's own cognitive state and interaction history, representing a vanilla agent-driven approach to peer participation. Further details are provided in Appendix~\ref{apx:expertEvaluationBaseline}.

\textit{Expert Participants.}
We recruited two expert mathematics teachers from different primary schools. E1 had 28 years of teaching experience, including 15 years teaching Grades~1--2; E2 had 29 years, including 5 years teaching Grades~1--2. 

\textit{Procedure.}
Using a custom replay interface (Figure~\ref{fig:expertReviewInterface}; Appendix~\ref{apx:expertEvaluationInstrument}), the experts independently and blindly rated the paired topics. After calibration, they reviewed the remaining sessions independently, and reconciled rating differences exceeding one point, while briefly explaining each inappropriate judgment to the researchers (Appendix~\ref{apx:expertEvaluationProcedure}). During independent review, the experts agreed exactly on 65.0\% of topics and differed by at most one point on 97.0\%; the linearly weighted Cohen's $\kappa$ was 0.466. Appendix~\ref{apx:expertEvaluationProcedure} provides a distribution-sensitive interpretation.

\textit{Measures.}
Experts rated the overall pedagogical appropriateness of each topic---considering both its instructional value and its fit with the learner and ongoing interaction---from 1 (\textit{very inappropriate}) to 5 (\textit{very appropriate}). A rating of 3 or below required a reason, provided through predefined tags or written feedback. After reconciliation, the two ratings were averaged to obtain the final message-level appropriateness score. For the reasons for inappropriateness, we retained all distinct reasons identified by either expert, interpreting them with reference to the experts' tags, comments, and consensus-meeting explanations.

\section{Findings}

Our findings are organized around the three research questions. We first characterize how children engage in and experience multimodal peer learning with \projectname{} (RQ1), then examine differences associated with peer composition (RQ2), and finally evaluate the pedagogical appropriateness of learning topics generated from children's interactions (RQ3). For readability, we report only $p$ values for statistical comparisons in the main text with full details in Appendix~\ref{apx:resultStatistics}.

\subsection{Children's Engagement and Experience in Multimodal Peer Learning (RQ1)}

Children engaged with \projectname{} in varied but predominantly learning-focused ways, and they generally reported positive experiences with the system; more than half also showed observable positive changes in their post-test artifacts. Their engagement extended beyond chat, as peer interactions elicited some children to create additional artifacts during learning. Conversational participation varied substantially across children and message types, with direct invitations and adaptive follow-ups associated with higher response rates.

\begin{figure}[htbp]
  \centering
  \includegraphics[width=0.97\linewidth]{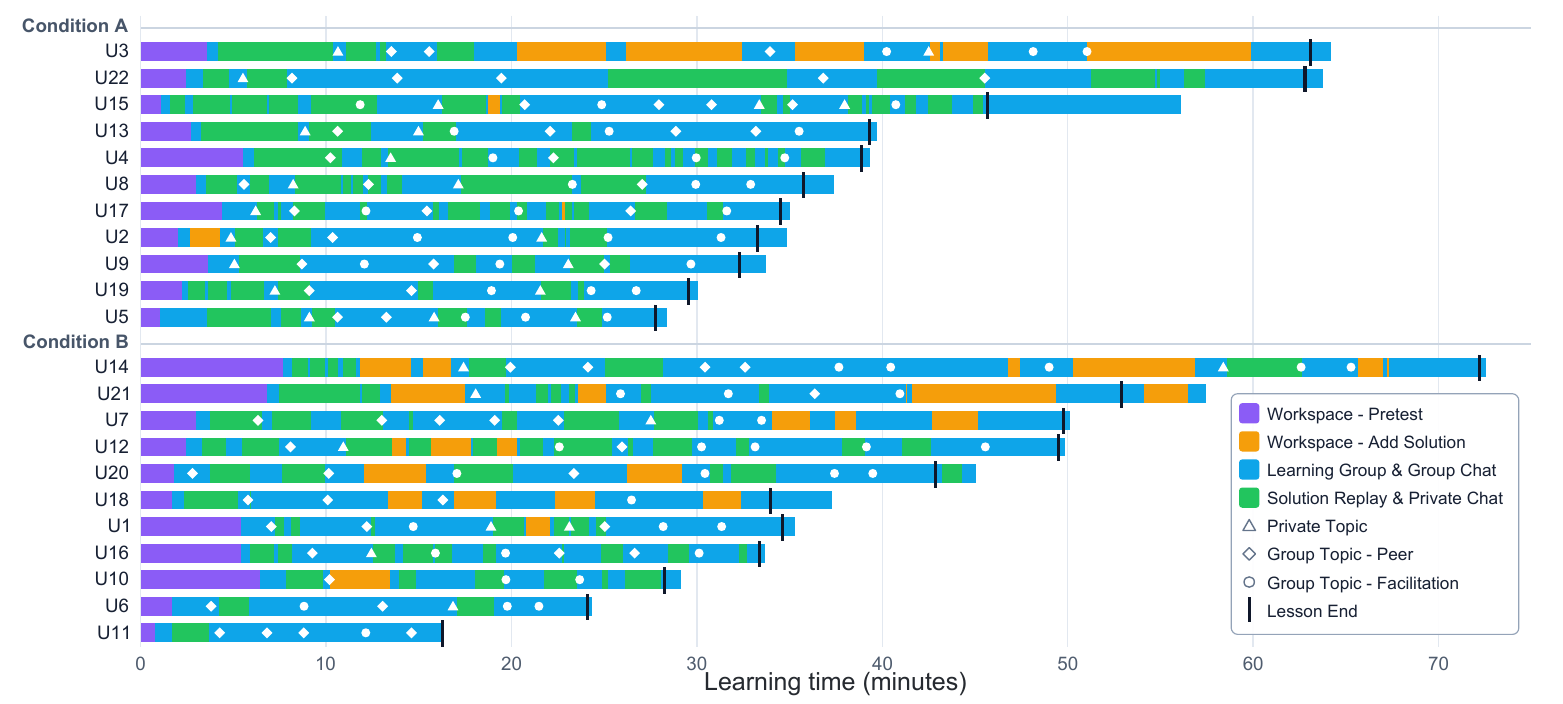}
  \caption{Interaction timelines for all 22 users, grouped by condition and ordered by total learning time. Colored segments show time spent in different views. White symbols mark different kinds of topics; vertical ticks mark lesson completion, after which users could continue interacting.}
  \Description{A horizontal stacked-timeline chart with one row per participant and an x-axis from 0 to about 75 minutes. Eleven Condition A participants appear above eleven Condition B participants, with rows ordered from longest to shortest session within each condition. Purple, orange, blue, and green segments encode the pre-test workspace, additional-artifact workspace, learning-group and group-chat view, and solution-replay and private-chat view. White triangles, diamonds, and circles mark private, peer-led group, and facilitation-led group topics. Black vertical ticks show lesson completion, and several rows contain colored activity segments after the tick.}
  \label{fig:6-timeline}
\end{figure}

\subsubsection{\textbf{\upshape Overall Engagement and Experience}}

All 22 children participated in the learning session, with timelines shown in Figure~\ref{fig:6-timeline}. One child (U11) chose to end it early, while the remaining 21 completed all topics orchestrated by the system. Sessions lasted $41.5 \pm 14.3$ minutes on average (range: 16.4--72.6), covering $7.1 \pm 1.9$ learning topics (range: 3--11). Children sent $17.5 \pm 16.1$ messages per session on average (range: 0--79). \textbf{Most messages remained focused on peer learning}: 73.8\% concerned mathematical learning content and 14.0\% concerned coordinating peer interactions, whereas only 12.2\% were off-task casual conversation.

Twenty children (except U11 and U22) subsequently completed the post-test, all producing correct answers. \textbf{Notably, 11 children (55\%) showed observable positive changes compared to their pre-test artifacts}, including newly adopted calculation strategies (4 children), new visual representations (8 children), or improved correctness (2 children); these categories were non-exclusive. U6, for example, attempted to represent subtraction with manipulatives in the pre-test but did not successfully show regrouping, while in the post-test, he correctly represented it using a strategy previously shown by a peer, later explaining that he was \Q{inspired by others}.

Overall, children exhibited high acceptance of \projectname{}. Compared with their usual mathematics learning experiences, 17 rated \projectname{} as more or somewhat more interesting. After finishing the session, some children spontaneously expressed strong interest in future use, asking if they could join this as an extracurricular class or download the application at home, or responding \Q{Great!} when they mistakenly thought another learning session was about to begin (U3, U19, U22). U12 expressed a desire to play with the peers in person and, after being told that they were virtual, immediately asked her mother to buy the entire setup.

\subsubsection{\textbf{\upshape Peer Interaction Elicited Further Artifact Creation}}

Peer interaction frequently elicited children to return to mathematical creation rather than remaining within the chat interface. During peer interaction, 11 children voluntarily created 29 additional artifacts ($M = 2.6 \pm 1.5$ among these children, range: 1--5).

These additional artifacts were often closely tied to ongoing peer interaction. We observed three primary contextual triggers: \textbf{(1) Peer sharing}: three children (U3, U20, U21) created and immediately shared artifacts right after seeing a peer share one in the group chat; \textbf{(2) Topic-prompted evaluation}: three (U7, U14, U21) authored new artifacts in response to the final \textit{Overall Evaluation} topic, which invited them to express what they considered a better approach; and \textbf{(3) Peer help-seeking}: three (U1, U12, U15) created artifacts to demonstrate a method while responding to a peer's request for help. For example, in response to repeated inquiries and scaffolding from a struggling peer, U12 consecutively authored three distinct artifacts, progressing from an equation to a visual demonstration and then to a step-by-step computational diagram. Social comparison could also motivate creation; as U3 remarked before entering the workspace, \Q{I feel like I'm falling behind everyone---I've only done two!}

The content of these new artifacts also reflected ongoing exposure to peers: 8 additional artifacts closely resembled a peer solution the child had previously viewed; 2 followed suggestions raised in chat; and 6 improved upon the child's own earlier work. Thus, peer artifacts functioned not only as objects for discussion, but also as resources for subsequent mathematical production. 

\subsubsection{\textbf{\upshape Participation Varied with Message Framing}}

Children's response rates varied substantially with conversational framing: explicit invitations and progressively structured follow-ups lowered the barrier to participation.

\begin{figure}[htbp]
  \centering
  \includegraphics[width=0.6\linewidth]{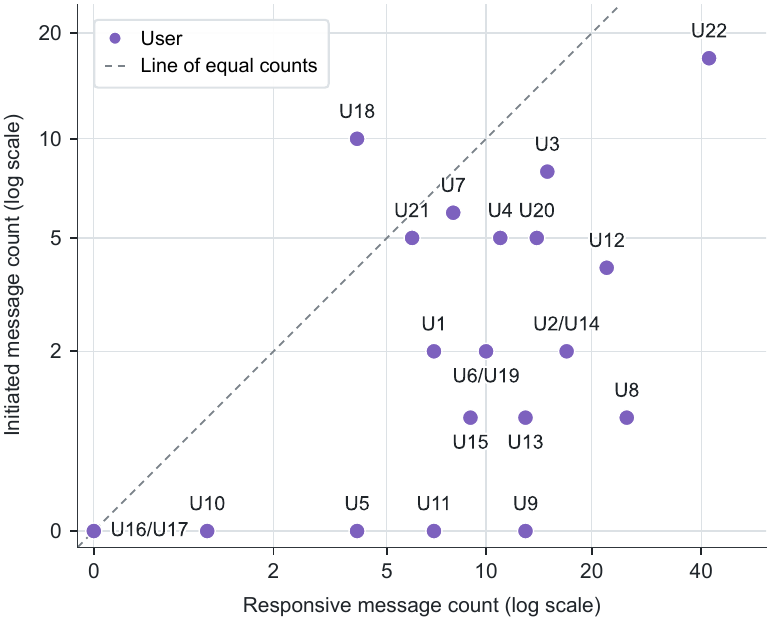}
  \caption{Learning-related message counts for each child, separated into user-initiated and responsive messages. Both axes use logarithmic scales, and the dashed diagonal denotes equal counts. Most children fall below the diagonal, indicating more responsive than user-initiated messages.}
  \Description{A scatterplot with responsive message count on the horizontal axis and initiated message count on the vertical axis, both shown on logarithmic scales. Twenty-two purple points are labeled by participant ID, and a dashed diagonal marks equal counts. Most points lie below the diagonal. U18 is the only clear point above it, with about 4 responsive and 10 initiated messages. U22 has the highest counts, with about 42 responsive and 18 initiated messages, while several participants have zero initiated messages.}
  \label{fig:6-messages}
\end{figure}

Overall, children exhibited diverse participation styles, ranging from passive observation to largely responsive, and to highly active, self-initiated engagement. As shown in Figure~\ref{fig:6-messages}, most children sent more responsive messages than self-initiated ones (with U18 being an exception). Children who responded rarely also produced almost no self-initiated messages, whereas initiative only began to emerge among highly responsive children.

\begin{table}[htbp]
  \centering
  \caption{Learner response rates by message framing. Open Q, Closed Q, Peer, All, and Learner refer to group chat; Private refers to private-chat questions.}
  \Description{A two-row table comparing response counts and rates across message forms and question targets. Questions received more responses than non-questions, 42.3 versus 18.9 percent; closed questions received more responses than open questions, 40.4 versus 27.6 percent. Response rates increased from questions directed to a peer, at 0 percent, to the whole group, at 28.9 percent, the learner, at 68.9 percent, and a learner in private chat, at 76.5 percent.}
  \label{tab:response-framing}
  \small
  \begin{tabular}{@{}lcccc@{}}
    \toprule
    \shortstack[l]{\textit{Message form} \\ \strut} &
    \shortstack{\textbf{Non-Q} \\ 52/275 (18.9\%)} &
    \shortstack{\textbf{Q} \\ 187/442 (42.3\%)} &
    \shortstack{\textbf{Open Q} \\ 48/174 (27.6\%)} &
    \shortstack{\textbf{Closed Q} \\ 74/183 (40.4\%)} \\
    \midrule
    \shortstack[l]{\textit{Question target} \\ \strut} &
    \shortstack{\textbf{Peer} \\ 0/19 (0.0\%)} &
    \shortstack{\textbf{All} \\ 80/277 (28.9\%)} &
    \shortstack{\textbf{Learner} \\ 42/61 (68.9\%)} &
    \shortstack{\textbf{Private} \\ 65/85 (76.5\%)} \\
    \bottomrule
  \end{tabular}
\end{table}

Across all 717 response-eligible incoming messages from peers and system facilitations, the overall response rate was 33.3\% and varied substantially with message characteristics (Table~\ref{tab:response-framing}). Questions were more likely to receive a response than non-question messages ($p<.001$). Response rates also varied markedly by target: within group chats, questions directed to another peer received no user replies, while questions broadcast to the whole group or directed at the learner received progressively more responses. Private-chat questions reached the highest response rate. Additionally, closed-ended questions in group chats were more likely to receive answers than open-ended questions ($p=.011$).

\begin{figure}[htbp]
  \centering
  \includegraphics[width=0.43\linewidth]{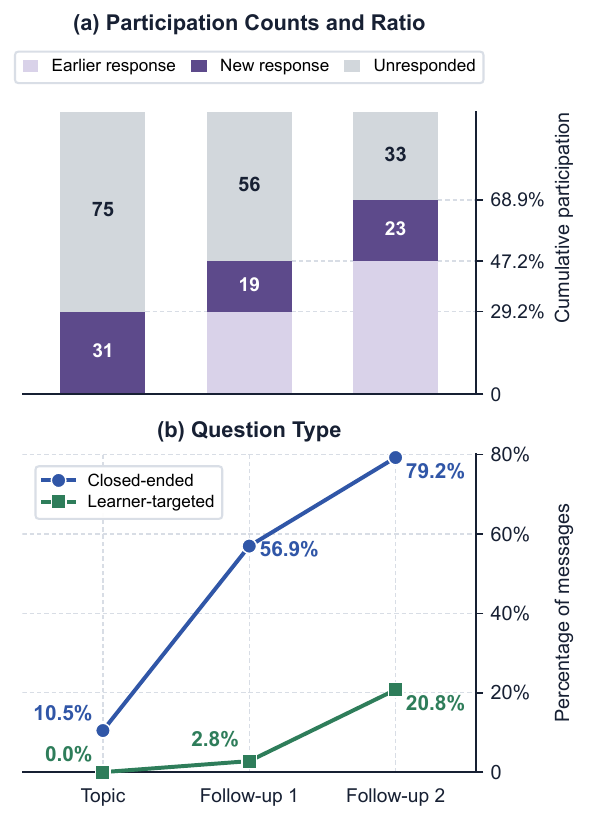}
  \caption{Participation across 106 group topics and their adaptive follow-ups. (a) Of the 106 topic messages, 31 received a direct response (29.2\%); the first and second follow-ups received 19 and 23 additional responses, increasing cumulative participation to 47.2\% and 68.9\%, respectively. (b) Across successive turns, messages increasingly used closed-ended questions and targeted the learner. These two message characteristics are not mutually exclusive.}
  \Description{A two-panel chart showing three successive turns: Topic, Follow-up 1, and Follow-up 2. Panel a contains three stacked bars for 106 topics. The initial topic bar shows 31 responses and 75 unanswered topics. Follow-up 1 adds 19 responses, leaving 56 unanswered; Follow-up 2 adds 23 responses, leaving 33 unanswered. The corresponding cumulative participation rates are 29.2, 47.2, and 68.9 percent. Panel b contains two line series. Closed-ended questions increase from 10.5 to 56.9 to 79.2 percent, while learner-targeted messages increase from 0 to 2.8 to 20.8 percent.}
  \label{fig:6-followups}
\end{figure}

Moreover, \textbf{adaptive follow-ups recovered participation} in many topics that did not initially receive a response. As shown in Figure~\ref{fig:6-followups}, among 106 eligible group topics, 29.2\% received an initial response. For those that did not, the system could issue up to two peer follow-ups; an additional 17.9\% received a response to the first follow-up and another 21.7\% to the second, resulting in a cumulative participation rate of 68.9\%. Across these three turns, the proportion of closed-ended questions increased sharply (10.5\% $\rightarrow$ 56.9\% $\rightarrow$ 79.2\%), as did explicit targeting of the learner (0.0\% $\rightarrow$ 2.8\% $\rightarrow$ 20.8\%). Together, this indicates that dynamically adjusting agency and progressively scaffolding questions from open-ended inquiry to targeted, low-barrier verification can re-engage children.

\subsection{Engagement and Experience across Peer Compositions (RQ2)}

Children's additional artifact creation, peer interaction, and perceptions of the group differed across the two peer compositions.


\subsubsection{\textbf{\upshape Leading Groups Elicited More Additional Artifact Creation}}

Children in Condition B (Leading Group) created significantly more additional artifacts than those in Condition A (Mixed Group) ($p=.037$). Descriptively, 8 children in Condition B created 22 additional artifacts, compared with 3 children and 7 artifacts in Condition A. This pattern is consistent with greater exposure to Leading Peers providing children with more material or motivation for further solution creation.

Interestingly, despite being surrounded exclusively by Leading Peers in Condition B, children did not perceive their peers as more mathematically capable, nor did they report greater peer pressure or perceived mathematical inferiority relative to their peers (Appendix~\ref{apx:resultStatistics}). Interview responses suggested that children inferred peer ability from varied and sometimes ambiguous cues, including the neatness and speed of peers' artifacts and how they interacted. For example, U12 interpreted continued discussion as a sign of lower ability: \Q{I could tell right away---they just kept discussing it.} We further discuss this pattern in \S~\ref{sec:discussion-pedagogy}.

\subsubsection{\textbf{\upshape Mixed Groups Afforded More Help-oriented Interaction}}

Condition A, by design, included Struggling Peers and therefore afforded more opportunities for private help-oriented interaction, including \textit{Help Requests} and \textit{Clarification Requests}. Accordingly, the system initiated significantly more private topics in Condition A ($p=.006$), and these sessions also contained significantly more private-chat messages ($p=.029$). Children in Condition A were furthermore significantly more willing to learn with the same peers again ($p=.015$). U15, for example, said she liked a particular peer because that peer had \Q{talked to me first.}

Seven children explicitly described enjoying teaching or helping peers, suggesting that help-seeking Struggling Peers could provide children with a valued contributory role in the group. Teaching could also prompt reflection on their own work. As U8 explained, \Q{At first, I didn't notice that I was wrong. I only realized it when I was teaching him.}

Together, these results suggest a complementary pattern across peer compositions: Leading Groups were associated with more continued solution creation, whereas Mixed Groups afforded richer one-to-one interaction and were associated with stronger willingness to continue learning with the same peers.

\subsection{Pedagogical Appropriateness of Generated Learning Topics (RQ3)}

Finally, an expert evaluation assessed the pedagogical appropriateness of topic-initiating messages from \projectname{}'s learner-centered orchestration and a persona-centered baseline.

\begin{figure}[htbp]
  \centering
  \includegraphics[width=0.9\linewidth]{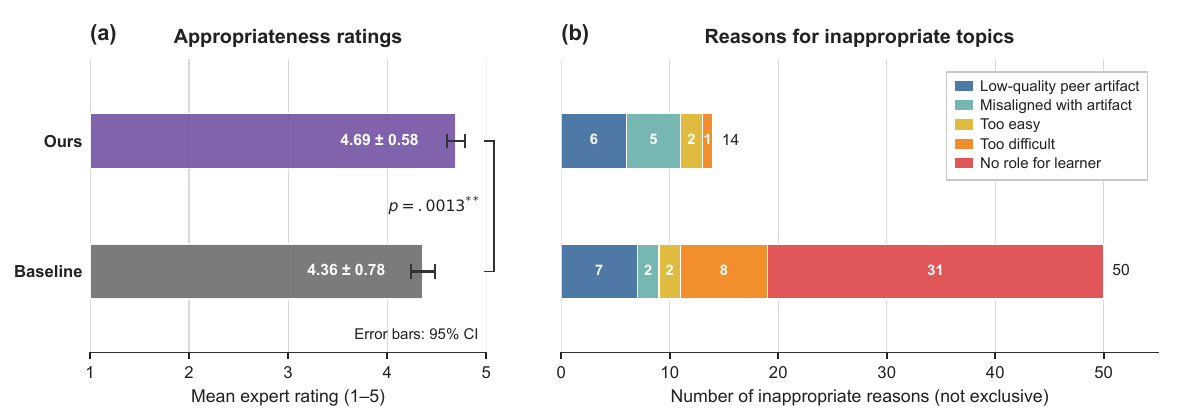}
  \caption{Post-hoc expert evaluation of learner-centered topic orchestration (Ours) and the persona-centered baseline across 157 topic rounds. (a) Mean expert ratings of pedagogical appropriateness on a five-point scale. (b) Counts of reasons coded by the experts for inappropriate topics; multiple reasons could be assigned to the same topic.}
  \Description{A two-panel horizontal bar chart. Panel A compares mean expert appropriateness ratings: Ours scores 4.69 plus or minus 0.58, and the baseline scores 4.36 plus or minus 0.78 on a five-point scale. Error bars show 95 percent confidence intervals, and a bracket reports p equals .0013. Panel B shows stacked counts of reasons for inappropriate topics. Ours has 6 low-quality-artifact, 5 artifact-misalignment, 2 too-easy, and 1 too-difficult reasons, totaling 14. The baseline has 7 low-quality-artifact, 2 artifact-misalignment, 2 too-easy, 8 too-difficult, and 31 no-role-for-learner reasons, totaling 50.}
  \label{fig:6-expert}
\end{figure}

Learner-centered topic messages achieved a high overall appropriateness score $4.69 \pm 0.58$, with 91.7\% regarded as appropriate (i.e., both experts rated $\geq 4$). Overall, learner-centered messages were rated significantly more appropriate than their persona-centered alternatives ($p=.0013$; Figure~\ref{fig:6-expert}a).

Figure~\ref{fig:6-expert}b summarizes the non-exclusive reasons for inappropriateness. Among the 13 \Q{inappropriate} learner-centered topics (i.e., at least one rating $<4$), the most common reasons concerned the underlying mathematical content: 6 were grounded in low-quality peer artifacts which experts considered inadequate as learning examples, and 5 contained misalignments with underlying artifacts, potentially misleading the learner. These cases primarily reflected limitations in solution quality or in the system's interpretation of multimodal mathematical artifacts.

In contrast, 43 persona-centered topics were rated as inappropriate, yielding 50 reason counts. Their dominant reason was interactional: 72.1\% (31/43) gave the learner little role in the topic, often because one virtual peer initiated a discussion primarily with another peer. Another common reason was excessive difficulty: 8 messages were considered too difficult for the learner because the topic was generated from the initiating peer's own perspective without accounting for the learner's current understanding. Thus, persona-centered topics were inappropriate predominantly because of learner involvement and cognitive alignment, whereas \projectname{}'s inappropriate topics more often reflected content quality and artifact interpretation.

Our analysis also identified valuable forms of persona-centered interaction. Experts explained that, when appropriately phrased, peer-to-peer discussions (e.g., one peer commenting on or asking another for clarification) could still capture the child's interest and support observational participation even if the learner did not respond directly.

\section{Discussion}

Drawing on our findings, we discuss two central questions for designing virtual peers in children's mathematics learning: how peer interaction takes shape, and how it becomes productive for learning. Beyond these, we discuss how artifact-mediated peer learning may extend to broader educational contexts.

\subsection{Supporting Peer Interaction through Multimodal Artifacts and Flexible Participation}

For young children, both the medium of peer interaction and the available modes of participation shape how accessible and meaningful that interaction becomes. Our findings show how multimodal solution artifacts can ground interaction in observable mathematical thinking, while flexible participation can accommodate children who prefer to initiate, respond, or observe. Children's emotional experiences also emerged as an important design consideration.

\subsubsection{\textbf{\upshape Multimodal Solution Artifacts as the Medium of Peer Interaction}}

Existing virtual peers are often designed first as \Q{conversational agents}, treating mathematical content merely as information exchanged through dialogue~\cite{yue2025MathVC,lyu2026Designinga}. Our findings suggest that for early mathematics learning, this relationship may be reversed: concrete mathematical solutions can themselves become the primary medium of interaction, with language revolving around these observable, authorable artifacts. Indeed, the act of authoring an artifact is, in itself, a form of communication~\cite{dillenbourg2006Sharing,hennessy2011Role}.

In \projectname{}, children did not merely observe peer solutions; they directed their inquiries and explanations at specific actions, symbols, and strategies embodied in those solutions. More importantly, exposure to peer methods frequently elicited children to imitate, adapt, or completely author new solutions, and then bring these new artifacts back into the interaction. Consequently, mathematical ideas can flow across artifacts through a continuous cycle of observation, explanation, imitation, modification, and sharing~\cite{stahl2014CSCL, koschmann1996CSCL}. Supporting multimodal mathematical peer learning is thus not simply a matter of adding whiteboards, images, or videos to a chat interface; it requires elevating mathematical artifacts to become the primary shared objects of social interaction.

This perspective also opens a broader design space. Children need not always start from a blank canvas. Future systems could allow them to directly annotate, remix, or build upon a peer's existing solution, thus making comparison and strategy transfer easier to enact. Likewise, virtual peers need not be restricted to statically pre-assigned solutions. As interaction unfolds, peers could also \Q{author} new artifacts, either through real-time AI generation or dynamic retrieval from a solution repository. Such capabilities could strengthen the \textit{symmetry of action} between learners and peers~\cite{dillenbourg1999What}, while further encouraging children to engage in mathematical construction and practice.

We therefore argue that the \Q{peer-like} quality of mathematical virtual peers need not rely only on conversational styles, personas, or anthropomorphic avatars~\cite{shamamyan2025CoDesigning}. It can also stem from a symmetry of mathematical agency: both children and peers can author, share, explain, and respond to one another's solutions. Such \emph{artifact-mediated peer interaction} turns mathematical representations from static learning resources into evolving objects of social interaction, grounding peer learning more directly in concrete mathematical thinking.

\subsubsection{\textbf{\upshape Flexible Participation across Initiating, Responding, and Observing}}

Our findings highlight multiple ways for children to participate in peer learning. Some frequently initiated questions and shared ideas, others primarily responded to explicit invitations, while still others preferred to observe. The critical design question, therefore, is whether the interaction lets children engage in ways that fit their preferences.

First, virtual peers can adjust the form of interaction to alter the threshold for participation. Our results showed that specific invitations directed toward the learner were more likely to receive responses than open, generalized group questions. When initial topics went unanswered, progressively direct and specific follow-ups brought some previously silent children into the discussion. Effective proactive peer behavior therefore depends on modulating the directness and openness of questions to provide accessible entry points. At the same time, observational participation can itself be a meaningful form of engagement~\cite{bandura1977Social,qian2024Learning}. For children who are reluctant to speak, observing questions and explanations exchanged among other peers may also support vicarious learning~\cite{tanprasert2023Scripted}.

Second, the timing of system-initiated interaction matters. While proactive peer messages can stimulate participation, they can also compete for children's attention and interrupt ongoing mathematical activity~\cite{wang2026Should}. For this reason, \projectname{} employs a strategy of implicit interaction~\cite{serim2019Explicating,chen2026When}: the system remains silent when the learner is engaged in active learning behaviors, and gradually takes greater initiative only after a sustained period of operational silence. This design also highlights a broader challenge: because operational silence is inherently ambiguous, learner-centered initiative should be grounded in continuous awareness of the learner's ongoing activity~\cite{liu2026Sensing}, rather than treating silence uniformly as disengagement.

This suggests that an ideal peer-learning system should personalize interaction through joint models of learners' mathematical cognition and social interaction styles~\cite{baxter2017Robot,ligthart2022Getting}. Systems could estimate a learner's initiative tendency, response habits, and topic preferences from prior interactions~\cite{schiaffino2004User}. For relatively passive children, a system might allow longer periods of observation, avoid repeated prompting, and offer clearer, lower-threshold invitations; for highly active children, it might reduce interventions and preserve more space for self-expression. In this way, different learners could first engage through modes that feel comfortable; over time, the system could gently nudge them toward more diverse forms of participation, helping them take fuller advantage of varied reciprocal learning opportunities, as we will discuss later.

Together, these findings position \emph{flexible participation} as a central design principle. Learner-centered virtual peers should adapt the threshold and timing of their interventions to each child's ongoing activity and interaction style.

\subsubsection{\textbf{\upshape Children's Emotional Experiences in Virtual Peer Learning}}

Because young children are still developing emotion regulation~\cite{schlesier2019Emotion,silkenbeumer2018Co}, virtual peer learning can introduce complex emotional dynamics alongside pedagogical opportunities. In our study, U9 frequently exhaled deeply, showed tension, and responded only passively; he relaxed only after the lesson ended, spending an extended time carefully inspecting the peers' artifacts. U11 communicated normally at first, but grew increasingly impatient as peer responses and audio continued to appear, eventually covering his ears, stating \Q{I don't want to listen anymore,} and asking to stop. U15 initially engaged actively in explaining and creating solutions, expressing delight when a peer used the same method. However, when that peer later praised a different approach, U15 sent a blaming message, became angry, and stopped responding across several consecutive topics. She rejoined the chat only after telling her peers that she was angry because \Q{they all said things that did not support her}.

These cases highlight the need for peer-learning systems to respond appropriately to children's potential emotional reactions. First, systems should enhance affect-awareness for young users through in-system interactions~\cite{arroyo2014Multimedia} and external signals such as tone of voice, facial expressions, gaze, and body movements~\cite{woolf2009Affectaware,kapoor2005Multimodal}, provided privacy is preserved~\cite{banzon2024Facial}. Second, systems should adapt the timing, frequency, and content of messages---for example, by enabling peers to affirm children's emotions~\cite{frivoldkostol2021Teachers}, perform relational repairs~\cite{dangol2026Toys}, or acknowledge the value in children's solutions. Together, affect-aware sensing and interactional adaptation can help young children remain safe and engaged in virtual peer learning.

\subsection{Turning Multi-Peer Interaction into Productive Mathematics Learning}
\label{sec:discussion-pedagogy}

Creating rich opportunities for peer interaction does not guarantee that learning will automatically occur. Our research highlights that translating interaction into learning depends on at least three interconnected factors: whether the system creates reciprocal learning oppurtunities, whether children can recognize the mathematical value of peers' methods, and whether interactions are orchestrated into a cognitively progressive learning process.

\subsubsection{\textbf{\upshape Enabling Reciprocal Learning through Heterogeneous Peers}}

The value of heterogeneous peers extends beyond presenting solutions at different levels; it lies in providing learners with cognitively complementary peers, thereby creating reciprocal learning opportunities~\cite{duran2005Styles,forman1998Exploring,kneser2001Collaboration}. Leading Peers can introduce novel strategies that push the learner's boundaries, whereas Struggling Peers can provide opportunities to explain, correct, and help. Crucially, these roles need not be rigidly fixed. A Struggling Peer may still possess a basic strategy that the learner has not used, while a Leading Peer may ask the learner for clarification or assistance on a particular issue.

Virtual peers can therefore support a dynamic form of \emph{reciprocal learning}. The direction of learning---\Q{who learns from whom}---can shift fluidly across different methods and topics. This differs from a typical teachable-agent paradigm, where the child is always cast as the tutor~\cite{chase2009Teachable,jin2024Teach}. Here, the contributory role emerges locally and temporarily within a holistic peer relationship of mutual learning, preserving the benefits of learning-by-teaching without permanently positioning participants as stronger or weaker.

This flexibility represents a practical advantage of virtual peers. Physical peer groupings are constrained by children's existing abilities and social relationships, whereas virtual groups can be tailored to an individual learner and adjust their roles as interaction unfolds. Heterogeneity can thus become a dynamic pedagogical resource within an enduring reciprocal relationship.

\subsubsection{\textbf{\upshape Making Virtual Peers' Expertise Recognizable to Children}}

Designing virtual peers with greater mathematical expertise does not ensure that children will recognize it. Although every peer in Condition B was designed as a Leading Peer, children did not perceive these peers as having higher ability. Their judgments of peer competence may not be based solely on the mathematical value of the artifacts, but also on social signals such as help-seeking behaviors, question frequency, and conversational style. For instance, some children interpreted peers' repeated pedagogical questions as a sign of ignorance, concluding that they themselves were the more knowledgeable learners.

This gap matters pedagogically: a novel method can influence learning only when the learner recognizes its value. This may be particularly challenging for younger children, who may rely on their own familiar methods and lack the metacognitive skills to actively compare and adopt new strategies~\cite{ishida2002Students,stern1992Spontaneous}. Systems should therefore construct cognitively useful peers while also making their expertise and learning value perceptible and interpretable to the child.

One possible direction is to leverage richer social and pedagogical signals to help learners identify content worth attending to~\cite{fusaro2008Children,lane2015Roles,tong2020Role}. This could involve having other peers explicitly endorse a valuable method, introducing lightweight facilitator guidance, or designing clearer expertise cues. Future designs could also integrate teacher-led instruction organically with peer learning, allowing teachers or facilitators to provide scaffolding at critical moments. Yet a strong authoritative presence may undermine the reciprocal and equal character of peer interaction. Balancing horizontal \emph{peer equality} with vertical \emph{pedagogical authority} remains a crucial challenge for future system design.

\subsubsection{\textbf{\upshape Orchestrating Peer Interaction along a Cognitive Progression}}

Pedagogical orchestration remains necessary for moving peer interaction toward deeper mathematical learning. Young children may not spontaneously move from understanding a method toward analyzing, comparing, and transferring it, nor identify common mathematical structures across multiple representations---all of which requires appropriate guidance~\cite{rittle-johnson2007Does,rittle-johnson2017Promoting,king1997ASK}. An underlying cognitive progression can guide open peer interaction from casual social exchange toward deeper cognitive activity.

This also explains the limitations of allowing virtual peers to initiate topics independently based on their own personas. As our expert evaluation demonstrated, persona-centered topic generation can produce naturalistic, socially plausible peer-to-peer messages, yet such messages frequently fail to provide a clear participatory role for the learner or align with the learner's current level of understanding.

Multi-peer learning systems therefore need to maintain both \emph{social openness} and \emph{pedagogical structure}. The interaction can feel open, equal, and emergent, while the system orchestrates topics along a cognitive trajectory, moving from observation and comprehension toward application, comparison, evaluation, and eventually, creation and transfer. Such orchestration can evolve from predefined sequences of learning activities~\cite{kollar2006Collaboration, vogel2017SocioCognitive} into a dynamic process that adapts to the learner's evolving understanding. This more adaptive form can preserve the openness of peer interaction while guiding children toward deeper conceptual understanding and strategy transfer.

\subsection{Extending Artifact-Mediated Peer Learning to Broader Educational Contexts}

Although this study focuses on children's mathematics, artifact-mediated peer learning may extend to broader educational contexts. In more advanced mathematics, learners still construct, interpret, and connect representations to understand abstract concepts\cite{fonger2019Meaningfulnessa}. Science learning likewise relies on dynamic representations that make abstract relationships observable. For example, \textit{Augmented Physics} turns static textbook diagrams into interactive simulations of topics such as optics, circuits, and kinematics~\cite{gunturu2024Augmented}. Such representations could become shared artifacts that learners and virtual peers manipulate and discuss together.

Scaling this paradigm, however, requires artifacts that are both credible and cognitively diverse. During development, we found that directly generating multimodal problem-solving processes with LLMs often produced overly polished and mature artifacts, missing the irregularities, partial errors, and subtle diversity of human reasoning. Similar limitations have been reported in LLM-based student simulation~\cite{wu2025Embracing}. We therefore assembled \projectname{}'s artifact repository from authentic child-produced artifacts and researcher-authored supplements. This approach was manageable for early mathematics, but constructing comparable repositories may become a bottleneck in domains where authentic artifacts are more costly to obtain or produce.

\section{Limitations and Future Work}

First, the generalizability of our findings is constrained by the small sample of 22 children aged 6--8, recruited from one Chinese city, and by our focus on one mathematics topic. The single, researcher-present laboratory session may also amplify novelty effects and cannot reveal how participation, peer relationships, or interest evolve over time. Longitudinal deployments across populations, topics, and everyday learning settings are needed.

Second, our immediate pre--post task and small between-subjects conditions provide limited evidence of learning effectiveness. Without a learning control condition or delayed measures, we cannot establish durable learning gains or transfer. The persona-centered baseline was evaluated offline by two experts and never entered children's interactions; an online comparison is required to understand its effects on participation and learning.

Finally, the pedagogical value of peer profiles and artifact assignments produced by our VLM analysis and hand-crafted rules remains unvalidated, while expert-identified low-quality artifacts call for stronger curation. We also held group size and total artifacts constant at four peers and approximately ten artifacts, leaving their effects on social presence and cognitive load unknown. Future work should validate and adapt peer construction, group size, and artifact exposure to support productive learning.

\paperacknowledgments

\bibliographystyle{config/ACM-Reference-Format}
\bibliography{references}

\appendix
\section{Formative Study}
\label{apx:formativeStudy}

\subsection{Mathematics Content and Tasks}
\label{apx:formativeProblems}

The formative study covered three mathematics topics, each at three progressively harder tiers, resulting in nine tasks (Table~\ref{tab:formativeProblems}).

\begin{table*}[htbp]
    \caption{Mathematics Content and Tasks Used in the Formative Study.}
    \Description{A two-column table listing nine formative-study tasks across three mathematics topics. Number-representation tasks ask children to represent 17, 46, and 358; addition-with-regrouping tasks ask them to calculate 6 plus 8, 15 plus 7, and 27 plus 14; and subtraction-with-regrouping tasks ask them to calculate 13 minus 7, 24 minus 15, and 45 minus 28.}
    \label{tab:formativeProblems}
    \centering
    \begin{tabular}{ll}
        \toprule
        \textbf{Content} & \textbf{Task} \\
        \midrule
        \multicolumn{2}{l}{\textbf{Number Representation}} \\
        Numbers within 20    & \Q{How can you represent the number 17?} \\
        Numbers within 100   & \Q{How can you represent the number 46?} \\
        Numbers within 1,000 & \Q{How can you represent the number 358?} \\
        \addlinespace
        \multicolumn{2}{l}{\textbf{Addition with Regrouping}} \\
        Addition within 20  & \Q{How can you calculate $6+8$?} \\
        Addition within 30  & \Q{How can you calculate $15+7$?} \\
        Addition within 100 & \Q{How can you calculate $27+14$?} \\
        \addlinespace
        \multicolumn{2}{l}{\textbf{Subtraction with Regrouping}} \\
        Subtraction within 20  & \Q{How can you calculate $13-7$?} \\
        Subtraction within 30  & \Q{How can you calculate $24-15$?} \\
        Subtraction within 100 & \Q{How can you calculate $45-28$?} \\
        \bottomrule
    \end{tabular}
\end{table*}

\subsection{Probe Implementation}
\label{apx:formativeProbe}

\paragraph{Peer-solution configurations.}
To expose participants to different forms of peer heterogeneity, we implemented three configurations for presenting peer solutions: (1) a developmental configuration that followed the developmental progression of the underlying mathematical strategies; (2) a diversity configuration that maximized variation among the presented artifacts; and (3) a researcher-selected configuration that enabled targeted presentation of unusual, incomplete, or flawed solutions. In the developmental configuration, the probe mapped the child's submitted solutions to a task-specific cognitive graph adapted from OpenCD~\cite{zheng2026OpenCD}, prioritized artifacts near the child's next developmental steps, and updated its estimate as artifacts were viewed. The diversity configuration represented solutions by their mathematical behavior features and selected artifacts dissimilar to both the child's solutions and previously presented artifacts. These two configurations initially displayed four artifacts and added two when the researcher requested more. In the researcher-selected configuration, flawed solutions were marked as needing help and opened with a short, pre-scripted request for assistance before the conversation began.

\paragraph{Virtual-peer interaction.}
The probe used one artifact-grounded Chat Agent based on Gemini 3 Flash Preview (temperature 0.7). For each turn, it received the associated task, the peer artifact's complete multimodal construction process and analysis summary, and the conversation history, and produced a concise reply from a peer persona aligned with the mathematical understanding reflected in that artifact. Children's speech was transcribed with Volcengine streaming ASR, and peer replies were streamed as speech using ElevenLabs' \texttt{eleven\_v3} model and a childlike voice. To further de-identify the original children's narration during artifact replay, its pitch was shifted.

\subsection{Child Participants}
\label{apx:formativeChildren}

Table~\ref{tab:formativeChildParticipants} presents the demographic details of the child participants.

\begin{table*}[htbp]
    \caption{Demographics of Child Participants in the Formative Study.}
    \Description{A table listing the ID, grade, gender, and age of eight formative-study participants. Four participants were in kindergarten and four were in Grade 1; four were female and four were male; ages ranged from 5 years 11 months to 7 years 5 months.}
    \label{tab:formativeChildParticipants}
    \centering
    \begin{tabular}{llll|llll}
        \toprule
        \textbf{ID} & \textbf{Grade} & \textbf{Gender} & \textbf{Age} & \textbf{ID} & \textbf{Grade} & \textbf{Gender} & \textbf{Age} \\
        \midrule
        S1 & K & Male   & 5y11m & S5 & 1 & Male   & 7y5m \\
        S2 & 1 & Male   & 7y4m  & S6 & 1 & Female & 7y2m \\
        S3 & K & Female & 6y5m  & S7 & K & Male   & 5y11m \\
        S4 & 1 & Female & 6y11m & S8 & K & Female & 6y4m \\
        \bottomrule
    \end{tabular}
\end{table*}

\paragraph{Task and configuration exposure.}
Table~\ref{tab:formativeProbeExposure} summarizes which configurations each child encountered for each task. We varied configurations to elicit reactions to contrasting peer sets within this exploratory probe.

\begin{table*}[htbp]
    \caption{Peer-solution configurations encountered by each child and task. D = developmental; V = diversity-maximizing; R = researcher-selected; -- = task not attempted.}
    \Description{A matrix showing which of three peer-solution configurations each of eight formative-study participants encountered on nine mathematics tasks. Columns group three number-representation tasks, three addition tasks, and three subtraction tasks; each attempted-task cell is marked developmental, diversity-maximizing, or researcher-selected, and unattempted tasks are marked with a dash.}
    \label{tab:formativeProbeExposure}
    \centering
    \small
    \begin{tabular}{lccccccccc}
        \toprule
        & \multicolumn{3}{c}{\textbf{Number Repr.}} & \multicolumn{3}{c}{\textbf{Addition}} & \multicolumn{3}{c}{\textbf{Subtraction}} \\
        \cmidrule(lr){2-4} \cmidrule(lr){5-7} \cmidrule(lr){8-10}
        \textbf{ID} & \textbf{17} & \textbf{46} & \textbf{358} & \textbf{$6+8$} & \textbf{$15+7$} & \textbf{$27+14$} & \textbf{$13-7$} & \textbf{$24-15$} & \textbf{$45-28$} \\
        \midrule
        S1 & R   & D & R & V & R & -- & -- & V & -- \\
        S2 & R   & -- & R & -- & -- & D & -- & -- & -- \\
        S3 & V   & R & -- & -- & R & D & -- & D & -- \\
        S4 & --  & V & R & -- & -- & D & -- & -- & D \\
        S5 & --  & V & R & -- & V & -- & -- & -- & V \\
        S6 & --  & V & R & -- & -- & -- & -- & -- & -- \\
        S7 & R & V & D & -- & V & -- & -- & -- & -- \\
        S8 & D   & R & -- & R & -- & -- & V & -- & -- \\
        \bottomrule
    \end{tabular}
\end{table*}

\subsection{Teacher Participants}
\label{apx:formativeTeachers}

The professional backgrounds of the teacher participants are detailed in Table~\ref{tab:formativeTeacherParticipants}.

\begin{table*}[htbp]
    \caption{Profiles of Teacher Participants in the Formative Study.}
    \Description{A five-row table describing the formative-study teachers by ID, total teaching experience, experience teaching Grades 1 and 2, gender, and school type. The three male and two female teachers had 7 to 17 years of total teaching experience and represented public schools, private schools, and a tutoring institution.}
    \label{tab:formativeTeacherParticipants}
    \centering
    \begin{tabular}{lllll}
        \toprule
        \textbf{ID} & \textbf{Teaching Exp. (years)} & \textbf{Grades 1--2 Exp. (years)} & \textbf{Gender} & \textbf{School Type} \\
        \midrule
        T1 & 9  & 4  & Male   & Public school \\
        T2 & 9  & 3  & Male   & Tutoring institution \\
        T3 & 14 & 7  & Female & Private school \\
        T4 & 17 & 17 & Female & Public school \\
        T5 & 7  & 5  & Male   & Private school \\
        \bottomrule
    \end{tabular}
\end{table*}

\section{System Design}
\label{apx:systemDesign}

\subsection{Artifact Analysis Agents}
Artifact preprocessing used the two VLM agents described in the main text. The Solution Artifact Comprehension Agent received the complete multimodal construction process and produced a concise description of the solution. The Behavior Detection and Cognitive Mapping Agent compared that process with the task-specific behavior rubrics and mapped the detected mathematical behaviors to competencies in the cognitive graph. When a behavior fell outside the predefined rubrics, this analysis included an additional outlier-diagnosis call to identify the relevant competencies. These outputs supported peer-profile construction and artifact assignment.

\subsection{Personalized Peer Construction}
\label{apx:personalizedPeerConstruction}

\paragraph{Peer-profile construction.}
The system used a topological enumeration algorithm over the task-specific cognitive graph to generate prerequisite-consistent states for Leading and Struggling Peers, and sampled the required number of each according to the group composition. Struggling Peers were also allowed to master selected foundational competencies that the learner had not yet demonstrated, creating opportunities for reciprocal tutoring.

\paragraph{Artifact assignment.}
Artifacts were assigned in three passes. First, the system covered the learner's mastery boundary, intermediate learning paths, and target competencies with artifacts compatible with each peer profile. Second, it added role-consistent incomplete or flawed artifacts to support correction and learning-by-teaching. Third, it supplemented the portfolios so that each peer had at least two artifacts, relevant terminal competencies were represented, and the group contained ten artifacts overall at least. When several assignments were possible, the priorities were candidate scarcity, cognitive fit, distributing repeated competency coverage across peers, balancing portfolio sizes, and novelty in mathematical behavior and modality.

\begin{figure}[htbp]
    \centering
    \begin{subfigure}{0.75\linewidth}
        \centering
        \includegraphics[width=\linewidth]{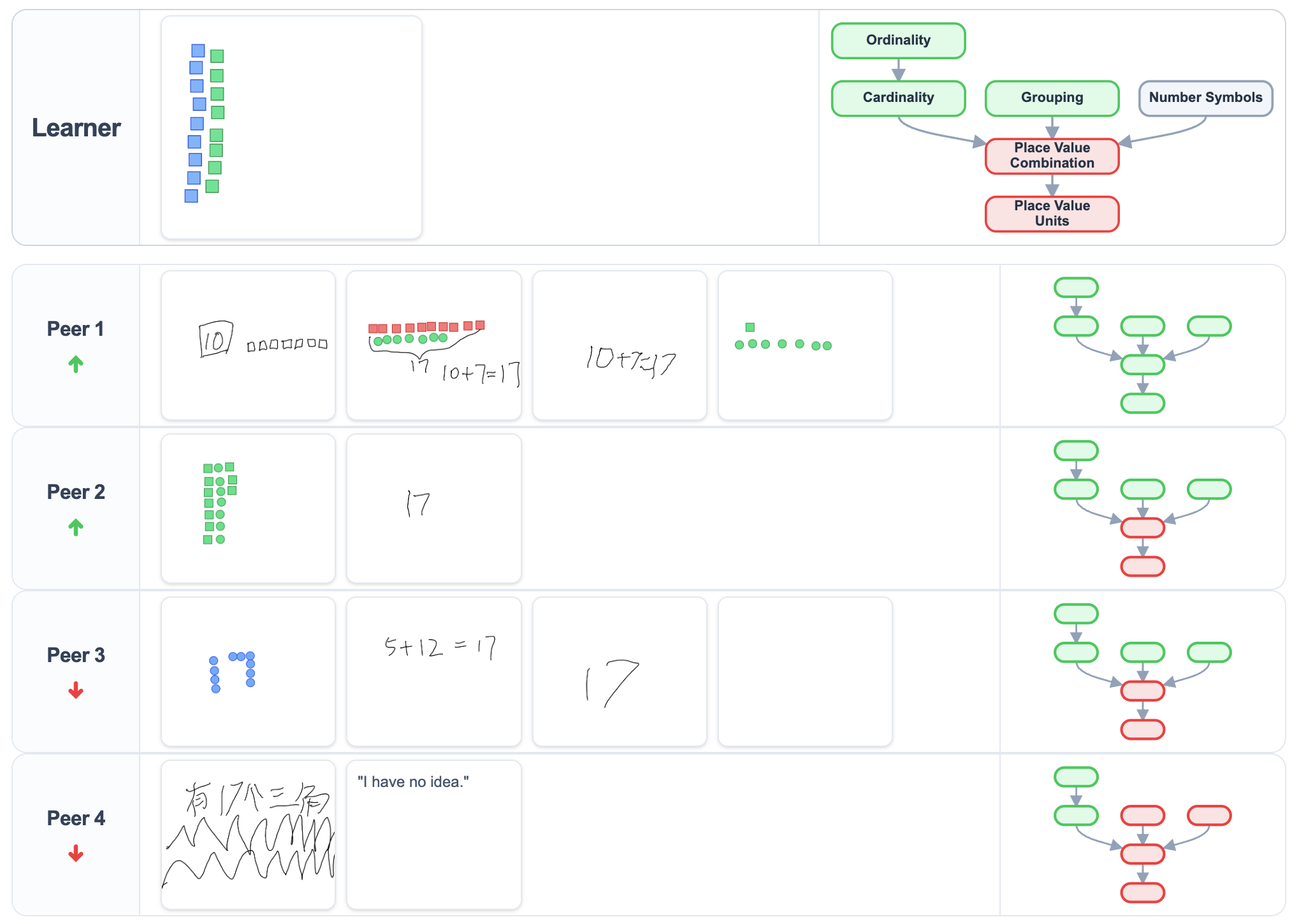}
        \caption{A Mixed Group with two Leading and two Struggling Peers for a task about representing 17.}
        \label{fig:apx-peer-construct-1}
    \end{subfigure}

    \begin{subfigure}{0.75\linewidth}
        \centering
        \includegraphics[width=\linewidth]{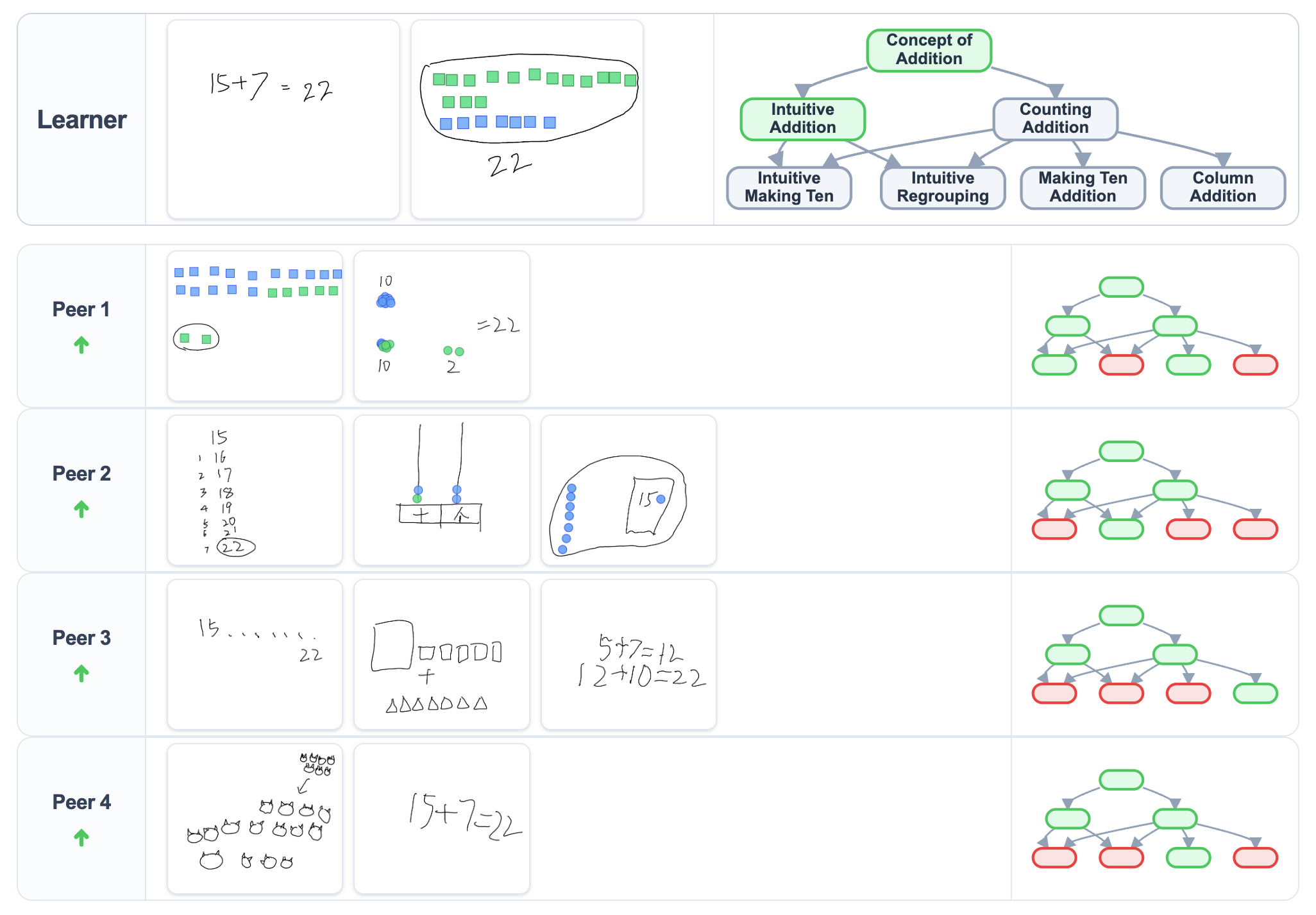}
        \caption{A Leading Group with four Leading Peers for the addition task $15+7$.}
        \label{fig:apx-peer-construct-2}
    \end{subfigure}
    \caption{Additional examples of personalized peer construction across tasks and group compositions.}
    \Description{Two vertically stacked examples, each organized as a learner row above four peer rows. In the first example, the learner represents 17 with two columns of colored squares. The learner's cognitive graph shows mastered ordinality, cardinality, and grouping, undemonstrated number symbols, and unmastered place-value combination and units. The generated Mixed Group contains two Leading Peers with complete grouping and equation artifacts and two Struggling Peers with less developed or flawed artifacts. In the second example, the learner solves 15 plus 7 using an equation and colored manipulatives. The generated Leading Group contains four Leading Peers whose portfolios present diverse strategies, including counting on, grouping, making ten, equations, and pictorial representations. In both examples, each peer row ends with a compact cognitive graph using green and red nodes.}
    \label{fig:apx-peer-construction}
\end{figure}

Figure~\ref{fig:apx-peer-construction} provides two additional construction examples: a Mixed Group for the task of representing 17 (Figure~\ref{fig:apx-peer-construct-1}) and a Leading Group for an addition task (Figure~\ref{fig:apx-peer-construct-2}).

\subsection{Adaptive Conversational Agency}
\label{apx:adaptiveConversationalAgency}

\paragraph{Activity intensity.}
To estimate interruption cost from interface behavior, the activity monitor grouped client events into four intensity levels. No-intensity events included opening or closing a chat panel and completing a replay; low-intensity events included navigating among peers, artifacts, and panels or pausing a replay; high-intensity events included listening to a message, playing or seeking within a solution replay, and changing a chat attachment; and extreme-intensity events included composing, recording, or sending a message and actively creating or submitting an artifact. These levels controlled how long the system continued to treat the learner as engaged: more intensive activity prolonged the waiting period, and extreme activity could cancel a pending proactive message.

\paragraph{Proactive follow-ups.}
During an ongoing group topic, sustained silence could trigger a \textit{pickup follow-up}, in which a peer re-entered the discussion with a more accessible invitation grounded in the current topic. Before the learner first responded, the system allowed up to two pickup follow-ups; after a response, it allowed one further pickup. If the learner continued to remain silent after the available pickups---or during a private topic, where pickups were not used---an \textit{end follow-up} produced a brief, concluding peer message without posing another question and closed the topic. Follow-ups were withheld while the relevant chat channel contained an unread message, preserving an opportunity for the learner to inspect the existing messages first.

\paragraph{Message-generation agents.}
The formal system used four message-generation agents, summarized in Table~\ref{tab:messageGenerationAgents}. The three chat agents used interaction-specific context and output constraints, while the Topic Agent followed the learner-centered orchestration described in the main text. All four used Gemini 3.6 Flash with a temperature of 0.

\begin{table*}[htbp]
    \caption{Message-generation agents used in the formal system.}
    \Description{A three-column table summarizing four message-generation agents, their triggers and inputs, and their outputs. The Group and Private Reactive Agents respond to learner messages in their respective chat channels; the Follow-up Agent responds to sustained silence during an ongoing topic; and the Topic Agent initiates a topic when the learner is idle and no topic is active.}
    \label{tab:messageGenerationAgents}
    \centering
    \small
    \begin{tabular}{p{0.19\linewidth} p{0.38\linewidth} p{0.35\linewidth}}
        \toprule
        \textbf{Agent} & \textbf{Trigger and Inputs} & \textbf{Outputs} \\
        \midrule
        Group Reactive Agent & A learner message in group chat; the current topic, peer profiles and artifacts, and group-chat history. & One or more peer replies with optional artifact attachments, together with whether to continue or close the topic. \\
        \addlinespace
        Private Reactive Agent & A learner message in private chat; the selected peer's profile and artifacts, the current topic, and relevant private- and group-chat context. & One or more replies from that peer with optional artifact attachments, together with whether to continue or close the topic. \\
        \addlinespace
        Follow-up Agent & Sustained silence during an ongoing topic; the interaction state, current topic, relevant channel history, and peer profiles and artifacts. & A peer message with optional artifact attachments, realized as either a pickup or end follow-up according to the interaction state. \\
        \addlinespace
        Topic Agent & An idle period with no ongoing topic; the task and cognitive graph, user and peer profiles and artifacts, prior topic history, and complete chat history. & The next topic type, channel, speaker or facilitation role, message, and referenced artifacts. \\
        \bottomrule
    \end{tabular}
\end{table*}

\section{User Study}
\label{apx:userStudy}

\subsection{Child Participants}
\label{apx:userStudyParticipants}

Before the study, parents completed a brief background questionnaire reporting the child's birth month, gender, grade, prior use of math-learning platforms, usage frequency of electronic devices and conversational AI, regular extracurricular mathematics-learning experiences, and current mathematics progress. Among the 22 formal-study participants, 15 used electronic devices daily, 5 weekly, and 2 rarely; 11 used conversational AI frequently, 7 occasionally, and 4 had never used it. 11 had used mathematics-learning platforms, 17 had participated in some form of regular extracurricular or home mathematics learning, and 7 were learning ahead of their completed grade.

Table~\ref{tab:userStudyParticipants} reports the formal-study participants. Grade denotes the school year completed before the study in summer break; K denotes the final kindergarten year. Task tiers correspond to Table~\ref{tab:userStudyProblems}.

\begin{table*}[htbp]
    \caption{Demographics and Study Assignments of Child Participants.}
    \Description{A 22-row participant table reporting ID, completed grade, gender, age, study condition, task tier, device-use frequency, conversational-AI-use frequency, prior mathematics-platform use, and whether the child was learning ahead of grade. Conditions A and B each included 11 children; ages ranged from 6 years to 8 years 2 months, and task tiers ranged from 1 to 3.}
    \label{tab:userStudyParticipants}
    \centering
    \scriptsize
    \setlength{\tabcolsep}{3pt}
    \begin{tabular}{llllllllll}
        \toprule
        \textbf{ID} & \textbf{Grade} & \textbf{Gender} & \textbf{Age} &
        \textbf{Cond.} & \textbf{\shortstack{Task \\ Tier}} &
        \textbf{\shortstack{Device \\ Use}} & \textbf{\shortstack{Conversational \\ AI Use}} &
        \textbf{\shortstack{Prior Math \\ Platform Use}} &
        \textbf{\shortstack{Ahead of \\ Grade}} \\
        \midrule
        U1  & 1 & Male   & 7y5m  & B & 3 & Daily  & Occasional & No  & No  \\
        U2  & 1 & Male   & 7y7m  & A & 3 & Rarely & Occasional & No  & Yes \\
        U3  & 1 & Male   & 7y8m  & A & 2 & Daily  & Frequent   & Yes & No  \\
        U4  & K & Female & 6y6m  & A & 1 & Daily  & Frequent   & Yes & No  \\
        U5  & 1 & Female & 7y10m & A & 3 & Rarely & Occasional & No  & No  \\
        U6  & 1 & Male   & 7y4m  & B & 3 & Daily  & Frequent   & Yes & Yes \\
        U7  & K & Female & 6y7m  & B & 1 & Daily  & Frequent   & Yes & No  \\
        U8  & K & Female & 6y9m  & A & 2 & Daily  & Occasional & No  & Yes \\
        U9  & K & Male   & 6y2m  & A & 1 & Daily  & Occasional & Yes & No  \\
        U10 & 1 & Male   & 8y2m  & B & 3 & Daily  & Frequent   & No  & Yes \\
        U11 & K & Male   & 6y2m  & B & 1 & Daily  & Frequent   & Yes & No  \\
        U12 & K & Female & 6y9m  & B & 1 & Weekly & Frequent   & Yes & Yes \\
        U13 & K & Male   & 6y0m  & A & 1 & Daily  & Never      & Yes & No  \\
        U14 & K & Male   & 6y0m  & B & 1 & Weekly & Never      & Yes & No  \\
        U15 & K & Female & 6y2m  & A & 1 & Daily  & Frequent   & No  & Yes \\
        U16 & 1 & Male   & 7y6m  & B & 2 & Daily  & Never      & No  & No  \\
        U17 & 1 & Female & 7y4m  & A & 3 & Weekly & Occasional & No  & No  \\
        U18 & 1 & Female & 7y9m  & B & 3 & Daily  & Frequent   & No  & No  \\
        U19 & 1 & Female & 7y7m  & A & 3 & Daily  & Frequent   & No  & No  \\
        U20 & K & Male   & 6y6m  & B & 2 & Daily  & Occasional & Yes & Yes \\
        U21 & 1 & Female & 7y4m  & B & 3 & Weekly & Occasional & No  & No  \\
        U22 & K & Female & 6y11m & A & 1 & Weekly & Frequent   & Yes & No  \\
        \bottomrule
    \end{tabular}
\end{table*}

\subsection{Mathematics Content and Tasks}
\label{apx:userStudyProblems}

The user study focused exclusively on \textit{Subtraction with Regrouping}. Each participant completed one of the three content levels shown in Table~\ref{tab:userStudyProblems}. The initial task was used for the pre-test and learning, while the corresponding quiz task was used for the post-test. Across the three pre-test and learning tasks, the artifact repository comprised 130 retained and preprocessed artifacts: 92 from the OpenCD dataset, 6 collected during the formative study, and 32 authored by the researchers to supplement strategies underrepresented among the child-contributed artifacts. Across U1--U22, 62 unique artifacts were assigned to at least one peer portfolio: 18 of 45 in Tier~1, 19 of 48 in Tier~2, and 25 of 37 in Tier~3.

\begin{table*}[htbp]
    \caption{Mathematics Content and Tasks Used in the User Study.}
    \Description{A three-row table pairing the pre-test and learning task with the post-test task for each subtraction tier. The pairs are 13 minus 7 and 15 minus 8 for subtraction within 20, 24 minus 15 and 26 minus 8 for subtraction within 30, and 45 minus 28 and 62 minus 37 for subtraction within 100.}
    \label{tab:userStudyProblems}
    \centering
    \begin{tabular}{lll}
        \toprule
        \textbf{Content} & \textbf{Pre-test and Learning Task} & \textbf{Post-test Task} \\
        \midrule
        Subtraction within 20  & \Q{How can you calculate $13-7$?}  & \Q{How can you calculate $15-8$?} \\
        Subtraction within 30  & \Q{How can you calculate $24-15$?} & \Q{How can you calculate $26-8$?} \\
        Subtraction within 100 & \Q{How can you calculate $45-28$?} & \Q{How can you calculate $62-37$?} \\
        \bottomrule
    \end{tabular}
\end{table*}

\subsection{Post-study Questionnaire and Interview}
\label{apx:userStudyInstruments}

The child-facing questionnaire used labeled five-point pictorial faces adapted from the Smileyometer~\cite{hall2016Five}; labels were customized to each construct while retaining an ordered negative-to-positive response. The researcher read or clarified items when needed without suggesting an answer.

\textit{Peer liking.}
Children rated each peer from strongly dislike to strongly like and then ranked the four peers from most to least liked.

\textit{Perceived mathematics level.}
Children rated each peer's mathematics level from very low to very high, ranked the four peers, and placed themselves within that ranking.

\textit{Social pressure.}
Children rated four statements: \Q{After seeing their solutions, I felt less capable than them}; \Q{After our discussion, I felt less capable than them}; \Q{During the discussion, I worried that my method was poor}; and \Q{During the discussion, I was reluctant to share my ideas}. Each item used five labeled responses from endorsement to the opposite position.

\textit{Experience.}
Children compared learning this way with regular classes on two five-point scales: much more relaxed to much more tiring, and much more interesting to much more boring.

The structured interview followed a common set of questions aligned with the questionnaire; the researcher used brief clarification prompts to elicit reasons or concrete examples. The questions were: (1) Did you learn any new methods, and where did they come from? (2) Which of your own and the peers' methods did you like most? (3) Did you help peers, and do you like teaching peers? (4) Did you ask peers for help, and do you like doing so? (5) How did this experience compare with the ways you usually learn mathematics? and (6) Which parts did you like or dislike?

\subsection{Analysis of Child-study Data}
\label{apx:userStudyAnalysis}

\paragraph{Message analysis.}
For learner messages, two researchers first discussed an initial set of coding dimensions and independently coded all messages from four children. They then reconciled their interpretations and finalized a codebook covering content domain (mathematical learning, peer-interaction coordination, or off-task conversation) and conversational role (initiating, responding, or continuing the learner's own prior message). Each response was additionally linked to the specific prior message it answered. One researcher applied the codebook to the remaining learner messages, after which the second researcher reviewed the complete coding.

For the larger set of non-learner messages, the researchers defined a categorical codebook covering learning relevance, question versus non-question, open- versus closed-ended questions, and the target of group-chat questions (another peer, the whole group, or the learner). Given the volume of these messages, their comparatively standardized and explicit system-generated language, and the direct binary or three-way classification criteria, researchers used GPT-5.6-sol to apply these codes before they reviewed the resulting classifications. Channel, topic, and follow-up sequence were obtained directly from system metadata. A group-chat topic was eligible for the follow-up response analysis when the learner had not yet contributed to that topic, or when the learner's first message was coded as a response linked to a message within that topic.

\paragraph{Artifact analysis.}
One researcher initially coded the pre- and post-test artifacts. Both researchers then jointly reviewed every artifact from the beginning and coded its representational forms, calculation strategies, and correctness based on the mathematical behavior features synthesized in OpenCD~\cite{zheng2026OpenCD}. Comparing each child's two artifacts yielded three non-exclusive forms of positive change: adopting a new calculation strategy, introducing a new visual representation, and improving correctness. For additional artifacts created during learning, we aligned camera video, screen recordings, and system logs to reconstruct the surrounding interaction. We coded typical events preceding creation, actions immediately following it, and content relationships to the child's own or peers' earlier artifacts and chat suggestions.

\paragraph{Interview analysis.}
Because the interviews followed a shared set of structured questions, we conducted question-guided qualitative content analysis: responses were organized by interview question and recurring explanations and examples were grouped within each question. Quotations reported in the paper were translated from Chinese into English.

\paragraph{Quantitative analysis.}
Log-derived behavioral measures and questionnaire responses were summarized descriptively. Between-condition comparisons used two-sided, tie-aware exact Mann--Whitney $U$ tests, with Cliff's $\delta$ as the effect size. Comparisons of response rates across message characteristics used two-sided, user-stratified exact permutation tests to account for messages nested within children. Full test specifications and results are reported in Appendix~\ref{apx:resultStatistics}.

\subsection{Post-hoc Expert Evaluation}
\label{apx:expertEvaluation}

\subsubsection{Evaluation Dataset and Persona-centered Baseline}
\label{apx:expertEvaluationBaseline}

At each intervention point, the learner-centered candidate was the message delivered during the user study. The persona-centered candidate was generated using the interaction state immediately before that intervention, including the task, available artifacts, and conversation history.

We implemented the persona-centered baseline as two sequential LLM calls. The first selected the peer most likely to initiate a topic based on the peers' cognitive profiles, available artifacts, recent participation, and the current conversational context. The second prompted the selected peer to generate a topic from its own perspective, jointly determining the message and attached artifacts from its profile and relevant chat history. Both calls used Gemini 3.6 Flash with a temperature of 0.

\subsubsection{Review Interface and Rating Instrument}
\label{apx:expertEvaluationInstrument}

\begin{figure*}[htbp]
    \centering
    \includegraphics[width=\textwidth]{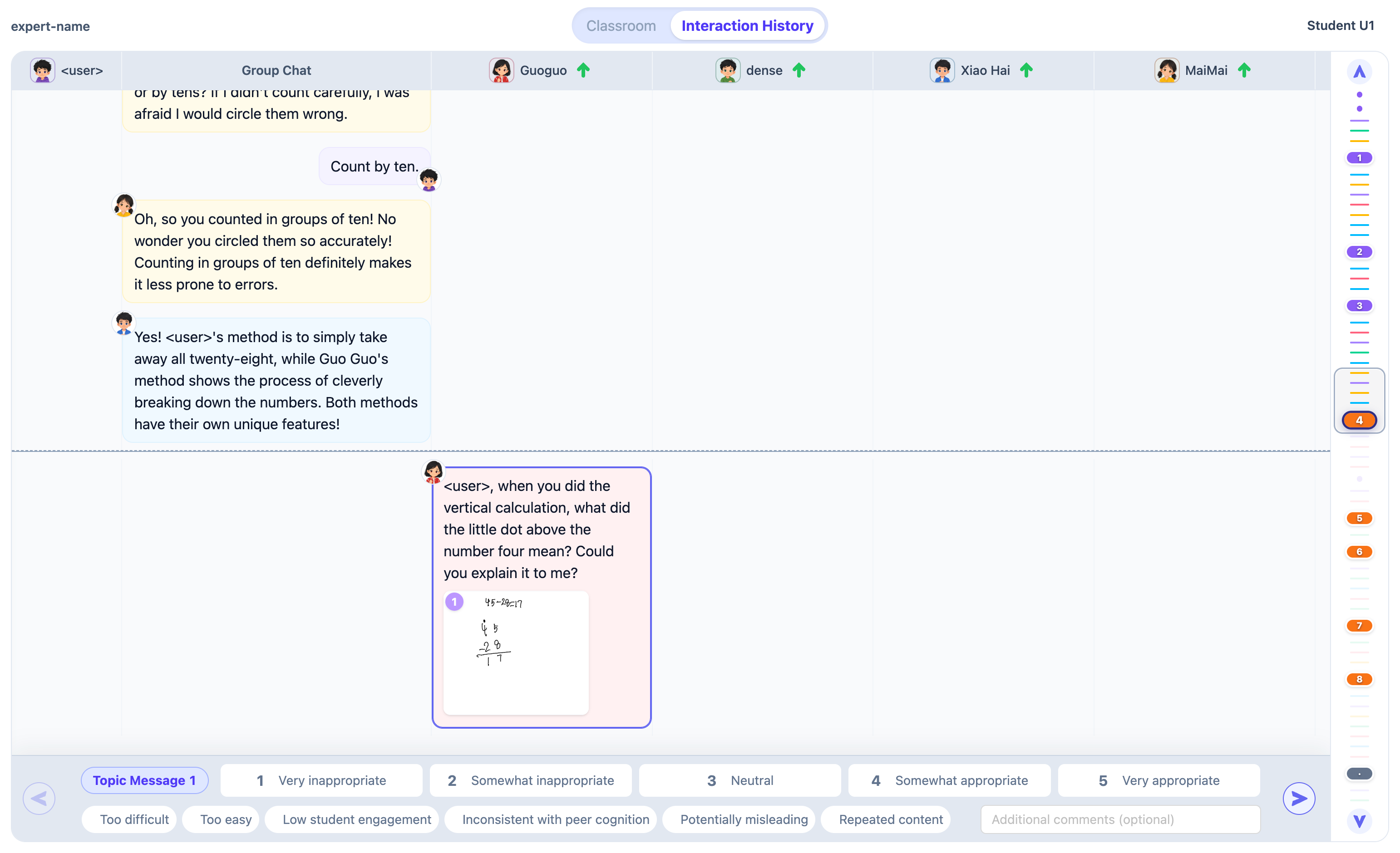}
    \caption{Expert-review interface. The interaction history preceding a topic appears above the divider, while one blinded candidate topic appears below it. Experts rated each candidate on a five-point appropriateness scale and could select reason tags or enter a comment.}
    \Description{A screenshot of the expert-review interface. The main area is divided into columns for the learner's artifacts, group chat, and private chats for four virtual peers. Prior conversation messages appear above a horizontal divider, and a candidate peer topic with an attached mathematics artifact appears below it. A vertical timeline on the right supports navigation through the session. Along the bottom are previous and next controls, a candidate-topic indicator, five appropriateness ratings from very inappropriate to very appropriate, six reason-tag buttons, and an optional comment field.}
    \label{fig:expertReviewInterface}
\end{figure*}

The custom interface (Figure~\ref{fig:expertReviewInterface}) replayed each session chronologically. At each intervention point, experts were provided with the conversation history up to that point. They were then presented with two candidate topic messages in a randomized order, with the condition identities hidden. The interface displayed each message's sender, channel, and attached artifacts. Experts independently rated both messages before the actual subsequent trajectory was revealed for the next turn.

Each topic was rated for its overall pedagogical appropriateness within the displayed context. The five-point scale ranged from 1 (\textit{very inappropriate}) to 5 (\textit{very appropriate}). For ratings of 3 or below, experts were required to select at least one predefined reason or provide a written explanation. The predefined tags included: too difficult for the learner, too easy for the learner, limited learner participation, inconsistent with the peer's cognitive profile, potentially misleading, and repetitive.

\subsubsection{Review Procedure}
\label{apx:expertEvaluationProcedure}

The review followed a three-phase protocol:

(1) \textit{Onboarding and Calibration.}
We first introduced the review task, rating scale, reason tags, and replay interface. The experts then independently reviewed 3 participant sessions. Subsequently, they met with the researchers to compare their judgments and discuss their interpretations of the scale and tags. The researchers clarified rating criteria and procedural questions, while the experts resolved discrepancies in the calibration cases. This established shared rating criteria for the remaining sessions.

(2) \textit{Independent Review.}
The remaining 19 sessions contained 133 intervention points, yielding 266 candidate messages. To ensure blind review, both experts rated the candidate messages before viewing the subsequent trajectory at each intervention point. The mean absolute difference between the two experts' pre-consensus ratings was 0.387. Because 85.5\% of the individual ratings were 4 or 5, the concentrated marginal distributions may partly explain the moderate weighted $\kappa$ despite high raw agreement.

(3) \textit{Consensus Meeting.}
After both independent reviews were completed, the experts used a comparison view showing their scores, selected tags, and comments side by side. They jointly revisited every candidate message for which their scores differed by more than one point and adjusted their ratings until the difference was no greater than one point. During this meeting, they also briefly explained to the researchers each message they regarded as inappropriate. The two resulting ratings were then averaged to obtain the final score for each message. For the reasons for inappropriateness, all distinct reasons identified by either expert were retained. The researchers interpreted and organized these reasons with reference to the experts' selected tags, written comments, and meeting explanations. The final analysis utilized these scores and reasons across all 22 participant sessions.

\section{Results}

\subsection{Statistical Details for Results Reported in the Main Text}
\label{apx:resultStatistics}

\paragraph{Peer pressure and perceived mathematical standing.}
Across the four peer-pressure items, the condition means ranged from 3.2 to 3.5 in Condition A ($SD=0.6$--$1.3$) and from 3.5 to 4.2 in Condition B ($SD=0.8$--$1.3$), with no significant differences ($U=41.5$--$52.0$, all $p\geq.149$). Children's mean ratings of their peers' mathematical levels were comparable (A: $4.4\pm0.6$, 3.3--5.0; B: $4.1\pm0.5$, 3.3--5.0), with no significant between-condition difference ($U_A=80.5$, $p=.191$). Children also did not rank themselves lower in Condition B (A: $2.6\pm1.3$, 1--5, $n=9$; B: $1.8\pm1.1$, 1--4, $n=11$; $U=66.5$, $p=.207$; lower ranks indicate higher perceived standing).
These questionnaire comparisons used two-sided, tie-aware exact Mann--Whitney $U$ tests with exhaustive permutation of the observed condition labels. Each comparison included 11 children per condition, except the self-ranking comparison, for which two responses in Condition A were unavailable ($n_A=9$, $n_B=11$).

\paragraph{Significant between-condition results.}

The following statistics correspond to the significant between-condition results reported in the first three Results subsections. Each comparison used a two-sided, tie-aware exact Mann--Whitney $U$ test, with the null distribution obtained by exhaustively permuting the condition labels across the 22 children ($n_A=n_B=11$; $\binom{22}{11}=705{,}432$ assignments). We report exact $p$ values. Cliff's $\delta$ is oriented as Condition A minus Condition B.

\begin{itemize}
    \item \textit{Additional solution creation.} Three children in Condition A created 7 additional solutions, whereas eight children in Condition B created 22. The per-child counts were $0.6\pm1.5$ (0--5) in Condition A and $2.0\pm1.7$ (0--4) in Condition B. The exact test yielded $U_A=31.0$, $p=.0370$, and Cliff's $\delta=-.488$.

    \item \textit{Private topics.} Sessions in Condition A contained 21 private topics in total, compared with 9 in Condition B. The per-child counts were $1.9\pm0.7$ (1--3) in Condition A and $0.8\pm0.8$ (0--2) in Condition B. The exact test yielded $U_A=101.5$, $p=.0060$, and Cliff's $\delta=.678$.

    \item \textit{Private-chat messages.} Sessions in Condition A contained 242 private-chat messages in total, compared with 79 in Condition B. The per-child counts were $22.0\pm21.6$ (2--71) in Condition A and $7.2\pm8.9$ (0--30) in Condition B. The exact test yielded $U_A=93.5$, $p=.0289$, and Cliff's $\delta=.545$.

    \item \textit{Willingness to learn with the same peers again.} Children in Condition A selected $3.8\pm0.6$ peers (2--4), compared with $2.5\pm1.6$ peers (0--4) in Condition B. The exact test yielded $U_A=94.0$, $p=.0152$, and Cliff's $\delta=.554$.
\end{itemize}

\paragraph{Expert evaluation of topic generation.}

\begin{itemize}
    \item \textit{Appropriateness ratings.} Across 157 paired topic rounds, the two experts' post-consensus ratings were averaged for each message. Learner-centered topics received $4.694\pm0.585$ points (median $=5.0$), compared with $4.360\pm0.784$ (median $=4.5$) for the persona-centered alternatives. A linear mixed-effects model of the paired rating difference, fitted by restricted maximum likelihood with a random intercept for child, estimated a learner-centered advantage of $\beta=0.338$ ($SE=0.091$), $t(21)=3.705$, $p=.0013$, 95\% CI $[0.148, 0.527]$.

\end{itemize}

\paragraph{Message characteristics and learner responses.}

The following comparisons used two-sided, user-stratified exact permutation tests. The procedure fixed each child's category-specific message counts and total number of responses, permuted category labels within child, and tested the response-rate difference. Both comparisons included 22 children, of whom 20 contributed nondegenerate permutation information. Wilson 95\% CIs are message-level descriptive intervals and do not account for within-child clustering.

\begin{itemize}
    \item \textit{Questions versus non-question messages.} Questions received responses in 187 of 442 eligible cases (42.3\%, 95\% CI $[37.8\%, 47.0\%]$), compared with 52 of 275 non-question messages (18.9\%, 95\% CI $[14.7\%, 24.0\%]$). The response-rate difference was 23.4 percentage points ($p=2.58\times10^{-13}$).

    \item \textit{Closed- versus open-ended questions in group chat.} Closed-ended questions received responses in 74 of 183 eligible cases (40.4\%, 95\% CI $[33.6\%, 47.7\%]$), compared with 48 of 174 open-ended questions (27.6\%, 95\% CI $[21.5\%, 34.7\%]$). The response-rate difference was 12.9 percentage points ($p=.0105$).
\end{itemize}


\end{document}